\documentclass[a4paper,fleqn,usenatbib]{mnras}

\usepackage{newtxtext,newtxmath}
\usepackage[T1]{fontenc}
\usepackage{ae,aecompl}

\usepackage{graphicx}	% Including figure files
\usepackage{epstopdf}

\newcommand{\gapprox}{\,\rlap{\lower 2.5pt % > ungefaehr =
\hbox{$\sim$}}\raise 1.5pt\hbox{$>$}\,}
\newcommand{\gsim}{\,\rlap{\lower 2.5pt % > ungefaehr =
\hbox{$\sim$}}\raise 1.5pt\hbox{$>$}\,}
\newcommand{\lapprox}{\,\rlap{\lower 2.5pt % < ungefaehr =
\hbox{$\sim$}}\raise 1.5pt\hbox{$<$}\,}
\newcommand{\lsim}{\,\rlap{\lower 2.5pt % < ungefaehr =
\hbox{$\sim$}}\raise 1.5pt\hbox{$<$}\,}

\def\eeq{\end{equation}}
\def\beq{\begin{equation}}

\title[Modelling spatially-resolved graveyards in galaxies]{Stellar  population modelling of neutron stars and black holes: spatially-resolved graveyards in MaNGA/SDSS-IV galaxies.}
\author[C. Maraston et al.]{
C. Maraston,$^{1}$\thanks{E-mail:claudia.maraston@port.ac.uk} 
M. Limongi,$^{2,4,5}$
J. Neumann,$^{3}$
L. Roberti,$^{2,5,7}$
A. Chieffi,$^{8,9,10}$
\and D. Thomas,$^{1,11}$
J. Lian$^{12}$
\\
$^{1}$Institute of Cosmology, University of Portsmouth, Burnaby Road, Portsmouth PO1 3FX, UK \\
$^{2}$Istituto Nazionale di Astrofisica - Osservatorio Astronomico di Roma, Via Frascati 33, I-00040, Monteporzio Catone, Italy \\
$^{3}$Max-Planck-Institut f{\"u}r Astronomie, K{\"o}nigstuhl 17, 69117 Heidelberg, Germany\\
$^{4}$Kavli Institute for the Physics and Mathematics of the Universe (WPI), The University of Tokyo Institutes for Advanced Study, Kashiwa, Chiba 277-8583, Japan \\
$^{5}$INFN. Sezione di Perugia, via A. Pascoli s/n, I-06125 Perugia, Italy \\
$^{6}$Konkoly Observatory, Research Centre for Astronomy and Earth Sciences, Konkoly Thege Mikl\'{o}s \'{u}t 15-17, H-1121 Budapest, Hungary \\
$^{7}$CSFK, MTA Centre of Excellence, Budapest, Konkoly Thege Miklós út 15-17, H-1121, Hungary \\
$^{8}$ Istituto Nazionale di Astrofisica—Istituto di Astrofisica e Planetologia Spaziali, Via Fosso del Cavaliere 100, I-00133, Roma, Italy \\
$^{9}$ Monash Centre for Astrophysics (MoCA), School of Mathematical Sciences, Monash University, Victoria 3800, Australia \\
$^{10}$ INFN. Sezione di Perugia, via A. Pascoli s/n, I-06125 Perugia, Italy \\
$^{11}$School of Mathematics and Physics, University of Portsmouth, Lion Gate Building, Portsmouth, PO1 3HF, UK \\
$^{12}$Yunnan University, Kunming, CN
}

\date{Accepted 2025 May 12. Received 2025 May 9; in original form 2024 March 28}

\pubyear{2025}

\begin{document}
\label{firstpage}
\pagerange{\pageref{firstpage}--\pageref{lastpage}}
\maketitle

\begin{abstract}
We update our stellar population models for the time evolution of the number and mass of massive remnants - neutron stars and black holes - with a new initial mass-remnant mass relation for core collapse supernovae. The calculations are based on hydrodynamical simulations and induced explosions of a subset of previously published pre-supernovae models spanning a wide range of stellar mass, metallicity and different values for rotation velocity. The resulting stellar population models predict lower numbers of neutron stars (by up to 0.3 dex) and higher numbers of black holes (by up to 0.8 dex), especially when stellar rotation is considered. The mass fraction locked in neutron stars and black holes is lowest in high-metallicity populations, with the largest number of remnants found at about half-solar metallicity. This mirrors the amount of available gas, ranging from 35 per cent to 45 per cent. We then apply our new models to IFU spectra for $\sim 10,000$~galaxies from the SDSS-IV/MaNGA survey for which we previously published spatially-resolved star formation histories. This allows us to probe spatially-resolved graveyards in galaxies of different types. The number and radial distribution of remnants depend on a galaxy's mass, star formation history and metal content. More massive and hence more metal-rich galaxies are found to host fewer remnants. Radial gradients in the number of remnants depend on galaxy mass mostly because of the mass-dependent profiles in mass density: the gradients are flat in low-mass galaxies, and negative in high-mass galaxies, particularly in Milky Way analogues.
%
%Models are available at \protect\url{www.icg.port.ac.uk/MaStar/}.
\end{abstract}


%%%%%%%%%%%%%%%%%%%%%%%%%%%%%%%%%%%%%%%%%%%%%%%%%%

%%%%%%%%%%%%%%%%% BODY OF PAPER %%%%%%%%%%%%%%%%%%
\section{Introduction}
\label{sec:intro}
Stellar populations - for a given initial mass function (IMF) - evolve on timescales determined by stellar evolution theory. As the newly-formed stellar population ages, stars will conclude their thermonuclear evolution on timescales inversely proportional to their stellar mass, leaving behind stellar remnants, such as black holes, neutron stars and white dwarfs. In absence of further star formation, the number and mass fraction of the most massive remnants will be set - at early epochs in dependance on the assumed mass threshold and remnant - and will not evolve any further. The number of white dwarfs on the other hand secularly increases with time after $\sim ~30$ Myr from the start, when the typical mass falls below  $\sim~8 M_{\odot}$, which is the limit above which stars go through supernovae and leave massive remnants such as neutron stars and black holes~\citep[e.g.][and references therein]{woosley_and_heger_2015}. The number and mass fraction of black hole and neutron star remnants is therefore established over a short period of time. In populations with extended stellar formation, additional remnants are added at later times, in dependence of the details of the star formation history. 

In our stellar population synthesis models \citep[][]{maraston_1998,maraston_2005}, stellar remnants for simple or complex star formation histories are calculated assuming the initial mass - final mass relation first published by \citet[][hereafter RC93]{renzini_and_ciotti_1993} and further amended in \citet[][]{maraston_1998}. Focusing on the part relevant to this paper, namely the most massive remnants such as NSs and BHs, the RC93 relation implies a constant remnant mass equal to the Chandrasekhar mass of $1.4 M_{\odot}$ for neutron star type of remnants, which in turn originate in the mass interval $8.5-40 M_{\odot}$. Above $40 M_{\odot}$, stars are assumed to leave a black hole remnant of half the initial stellar mass. Both relations are assumed to hold at any chemical composition of the stellar population, with metallicity effects being limited to small changes to the turnoff mass, hence to the relation between population age and the mass limits described above. 

In this paper we update our stellar population models by using a state-of-art initial mass - remnant mass relation based on 1D hydrodynamical sumulations in the framework of the thermal bomb. This is obtained by means of the HYPERION code \citep[][]{limongi_and_chieffi_2020}, applied to a database of progenitor models in the mass range $\rm 13-120~M_\odot$, with initial metallicities [Fe/H] = 0.3, 0, -1, -2, -3 and initial rotation velocities $ v_{\rm rot}=0,~150,~300{\rm km~s^{-1}}$ \citep[][]{limongi_and_chieffi_2019,RLC24}. The advantage of this relation is that it is the only one available that covers a wide range of stellar initial parameters, namely, initial mass, metallicity and rotation velocity. The same pre-supernovae models are used in \citet[][]{mapelli_etal_2020} to probe the effect of rotation on the formation of black holes. Differently from our work, however, these authors estimate the remnant mass by means of a criterion based on the compactness parameter of the core (see, e.g., their equation 2) at the time of the pre-supernova stage, without using the output of hydrodynamical calculations. In our paper, instead, the remnant masses have been obtained by means of hydrodynamical calculations, as described above (see Section~\ref{sec:imrm} for a detailed description). 
Moreover, our work also includes the calculation of NS remnants. Other relevant differences between this updated initial mass - remnant mass relation and the one provided by \citet{limongi_and_chieffi_2019} is that the database of progenitor stars is increased by adding one set of super-solar metallicity models with initial metallicity [Fe/H]=0.3. This is needed to model massive galaxies (e.g. Thomas et al. 2010). Moreover, the explosions are now performed with an updated version of the hydrocode and tuning the initial injected energy by requiring that the kinetic energy of the ejecta is $\rm 10^{51}~erg$ for all models. Note that all calculations are performed for single stars. A comprehensive presentation of our adopted initial mass - remnant mass relation and its  comparison with the literature is given in Section~2.

We then calculate stellar population models of remnants and their time evolution using the new initial mass - remnant mass relations and compare them with our standard output. We make our results available to allow comparison with results based on our standard output and to enable the build-up of more complex functions where stellar binarity is included. A comparison to other published relations \citep[e.g.][]{spera_mapelli_bressan_2015,mapelli_etal_2020} is found in Section~\ref{sec:discussion}. 

As an original application, we calculate models for the remnant population in real galaxies using the MaNGA \citep[Mapping Nearby Galaxies at Apache Point Observatory survey, ][]{bundy_etal_2016} project part of the Sloan Digital Sky Survey-IV \citep[SDSS-IV;][]{blanton_etal_2017}, containing in its final release \citep[][]{abdurro_etal_2017} Integral Field Units (IFU) spectra for 10,010 galaxies in the local universe. 
We calculate remnants using our previously published star formation histories for the galaxies (FIREFLY Value Added Catalogue (VAC) (Neumann et al. 2022\footnote{Available at https://www.sdss.org/dr17/manga/manga-data/manga-firefly-value-added-catalog}), by computing remnant mass and numbers for the individual single bursts models of given age and metallicity composing the galaxy star formation history. Therefore the dark remnant population is linked to the light emitted by living stars (the bright population), without any ad-hoc assumptions on the stellar population composition of the galaxies. mass-to-light ratio, etc. Moreover, as MaNGA is an IFU survey, we obtain - for the first time - spatially-resolved 'graveyards' of stars in galaxies of various morphological type, star formation history, age, chemical composition and stellar mass.

This paper is organised as follows. In Section~\ref{sec:canrel} we recapitulate the canonical relations and the population models based on these, while in Section~\ref{sec:newrel} we describe the new stellar evolution models for massive stars and introduce the calculations that were made specifically for this work. The description of how we implement the new relations in evolutionary population synthesis is in Section~\ref{sec:calculations}, while results are presented in Section~\ref{sec:results}. Section~\ref{sec:discussion} provides a discussion on the findings of the theoretical part, while Section~\ref{sec:manga} describe the application to galaxy data, namely the spatially resolved graveyards in galaxies. A general summary in Section~8 concludes the paper.   
\section{Canonical relations}
\label{sec:canrel}
Our canonical calculations of the mass budget of evolving stellar population models are presented in \citet[hereafter M98 and M05]{maraston_1998,maraston_2005}, where the full evolutionary population synthesis model is described and in \citet[][]{courteau_etal_2014}, where the link is set to the calculation of galaxy masses. Here we shall only recapitulate the most relevant points for the present work. 

The basis for calculating the stellar mass and number budget of stellar population models are: i) the stellar evolution theory adopted for the evolutionary population synthesis; ii) the initial mass - remnant mass relation. As for the first input, our standard stellar evolution prescription consists of isochrones and stellar tracks by
 \citet{cassisi_etal_1997} for ages larger than $\sim~30$~Myr and by \citet{genevatracks} for younger populations\footnote {Sets of models are also computed with 'Padova' stellar evolutionary models \citep{girardi_etal_2000}}. The latter ones are relevant to the present paper focusing on massive remnants, and what matters in particular is the turnoff mass, i.e. the mass corresponding to the exhaustion of the core-Hydrogen burning, as a function of age and chemical composition.

\begin{figure}
\includegraphics[width=\linewidth]{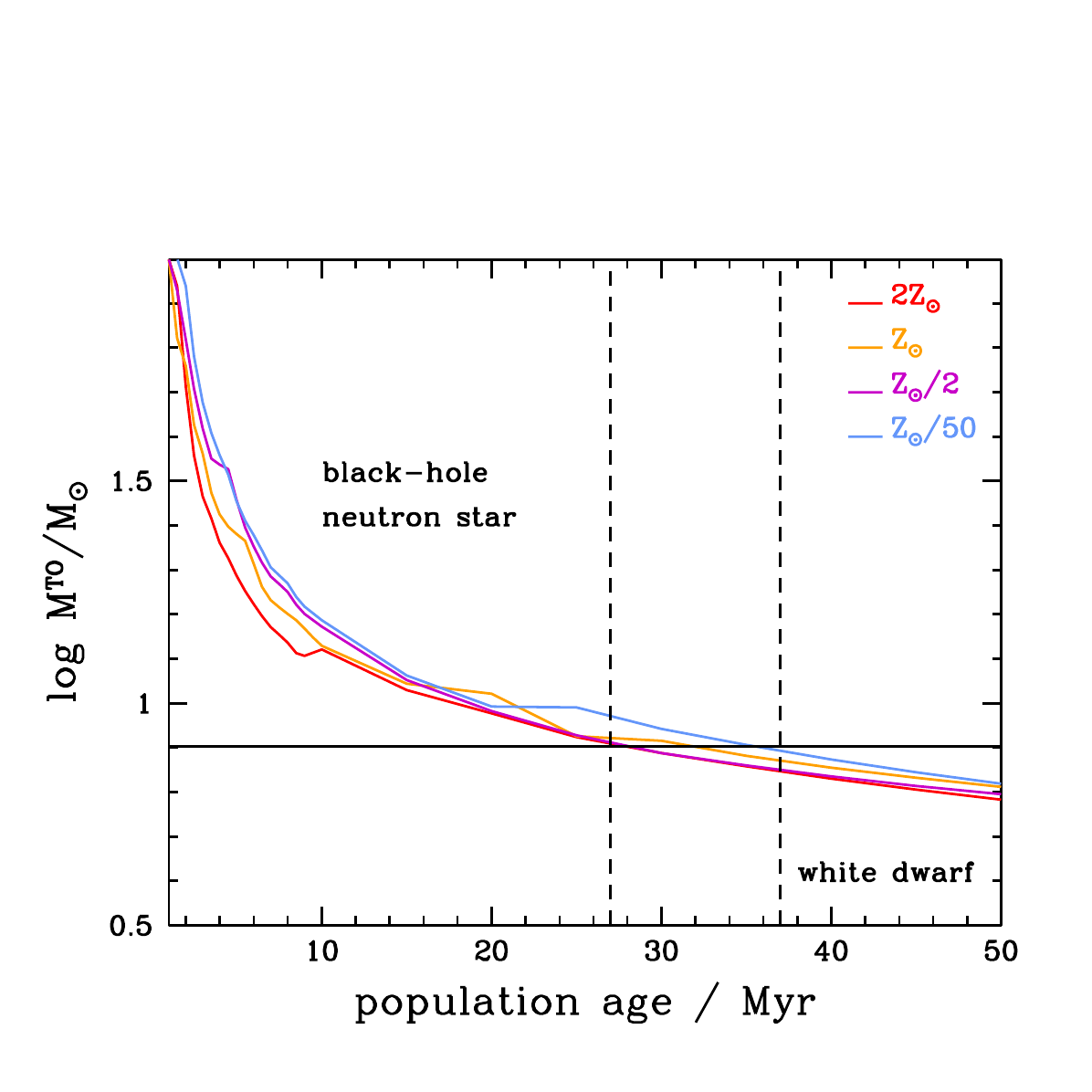}
\caption{Dependence on chemical composition (coloured labels) of the turnoff mass (y-axis) setting the evolutionary timescales (x-axis) for massive remnant production of single stellar generations. The solid line corresponds to $M_{TO}=8.5 M_{\odot}$, below which stars evolve into white dwarfs and the massive remnant production is halted. Dashed lines mark the time when this happens, as a function of the explored metallicity range.}
  \label{fig:mto}
\end{figure}

Figure~\ref{fig:mto} shows the turnoff mass vs the stellar population age, for various chemical compositions (coloured labels), ranging from a metal-rich population as typical of massive elliptical galaxies (red line) to a metal-poor population (blue line) as proper to globular clusters, dwarf galaxies or populations in the halos of massive galaxies. The plot focuses on the early evolutionary times when the production of massive remnants from a stellar generation, i.e. neutron stars and black holes, occurs. 

As can be seen, the effect of metallicity is mild, with higher turnoff masses generally corresponding to lower metallicities at a given time. Over this range in chemical composition the cutoff turnoff mass of $8.5 M_{\odot}$~below which stars no longer leave massive remnants and evolve into white dwarfs (solid line) happens between $\sim 27$ Myr and $\sim 37$ Myr (dashed vertical lines), for the adopted set of Geneva stellar tracks. Therefore for each single burst composing the star formation history of a galaxy, the massive remnant production is frozen after $\sim 40$ Myr since the burst. 

The second piece of information regards the remnant mass to initial mass relation. Our canonical calculations assume the initial mass - remnant mass relations from Renzini \& Ciotti (1993, hereafter RC93). Focusing on the most massive remnants, the RC93 relation foresees a neutron star remnant with mass equal to the Chandrasekhar mass of $1.4 M_{\odot}$\footnote{Observed neutron star masses are usually found to be a bit lower, in the 1.28-1.33~$M_{\odot}$~range, with recycled NS being consistent with 1.48~$M_{\odot}$ \citep[][]{oezel_etal_2012}. These differences have a negligible impact on the $M^{*}/L$~ratio of stellar populations.} as the end point of stellar masses between $8$ and $40 M_{\odot}$. These masses correspond to timescales of $\sim 30$~Myr and $\sim 3$~Myr in the evolutionary timescales of stellar populations, where exact values depend on metallicity (and the adopted stellar tracks) as described in Figure 1. Black hole remnants originate from the evolution of stars more massive then $40 M_{\odot}$, i.e. at ages younger than $\sim 3$~Myr. The black hole remnant weighs half the initial stellar mass. These relations are assumed to hold at any arbitrary chemical composition, the effect of metallicity just resulting in mild changes to the turnoff mass as described above. In the next section we shall describe the relations for stellar mass evolution we explore in this paper.
\section{The initial mass - remnant mass relation}
\label{sec:newrel}
\subsection{Preamble and state of the art.}
\label{sec:preamble}
The relationship between the initial mass of a massive star and the mass of its compact remnant after a core-collapse supernova (CCSN) explosion (the initial mass–remnant mass IM-RM relation) is one of the most fundamental and still evolving aspects of supernova research. This relation has profound implications for our understanding of stellar evolution, the formation of compact objects like neutron stars and black holes, and the explosion mechanisms of massive stars. Determining this relation is crucial not only for predicting the outcomes of supernovae, but also for interpreting a wide range of astrophysical observations, such as the light curves of supernovae, and the chemical enrichment of galaxies. The number and mass fraction of compact remnants affects the stellar mass-to-light ratio ($M^{*}/L$) of stellar generations \citep{maraston_1998} and galaxy dynamics (for a review see \citet{courteau_etal_2014}). Furthermore, particularly since gravitational waves have been experimentally confirmed \citep[][]{abbott_etal_2016}, the details of remnant calculations \citep{fryer_etal_2012} play a stronger role in astrophysics, as they impact on the interpretation of the detections and allow predictions for future targeted surveys such as LISA.

In spite of the importance of this relation, no first-principles IM-RM relation is currently available. The main reason is that the immense computational costs and difficulties in modeling the conversion of the energy carried by neutrinos leaking out of the neutrinosphere into thermal energy, i.e. the energy that ultimately drives the explosion of the star, have not allowed a clear-cut answer to the fundamental question, namely what is the remnant mass of a given progenitor star after the explosion.

The mass of the compact remnant, whether a neutron star or a black hole, depends on several factors: the mass and the internal structure of the progenitor star, the dynamics of the supernova explosion, and the interplay between various physical processes, including neutrino heating, shock dynamics, and fallback. A central aspect of understanding the IM-RM relation lies in determining how these factors scale with the initial mass of the progenitor and how they interact during the explosion. The IM-RM relation thus provides a framework that connects the properties of the progenitor star to the final outcome of the supernova explosion, shaping the final fate of the star.  

State-of-the-art 3D simulations, incorporating the most sophisticated treatment of neutrino transport, are able to follow the explosion only for a few seconds. This time is long enough to understand whether the explosion is successful or not, 
but it is too short to accurately determine the mass of the remnant since most of the fallback occurs over longer times \citep[see, e.g.,][]{Blondin_2003,Janka_1996, Burrows1995, Suwa2010, Suwa2013, Muller2012, Muller2015, Summa2016, Bruenn2016, Lentz_2012, Lentz2015, Burrows2019, Burrows2020, Mezzacappa2020, Bollig_2021, Wang2020, Wang2022, Burrows_2023, burrows2024theoryneutronstarblack, Vartanyan2023, Mezzacappa_2023, Powell_2023, Kinugawa2021, kinugawa2023, nakamura2024, Shibagaki_2024}.

Recently more self consistent 1D simulations have been presented based on a "calibrated" efficiency of the neutrino energy deposition and transport \citep{Ugliano:2012fvp, Boccioli+2023}. These simulations, however, were mainly focused on the study of the explodability of massive stars and did not present an IM-RM relation.

To overcome computational challenges, a simplified method for studying the explosion of a massive star involves artificially triggering the SN explosion by injecting energy at a chosen mass coordinate in the progenitor model. This energy input induces a shock wave, whose propagation through the stellar mantle is tracked using hydrodynamic simulations. Different schemes for energy injection have been employed, including the “thermal bomb” \citep{Thielemann+90}, the “piston model” \citep{Woosley:1995ip} and the “kinetic bomb” \citep{Limongi:2003ui}. Regardless of the method, the injected energy is in general tuned in order to obtain a given final kinetic energy of the ejecta or a given amount of ejected $\rm ^{56}Ni$. The advantage of these simplified approaches lie in their lower computational cost, allowing the simulations to follow the evolution of the shock over extended timescales. This makes it possible to estimate the fallback, the chemical composition of the ejecta, after the explosive nucleosynthesis is taken into account, and the supernova light curve. These kinds of 1D simulations have been extensively used in the literature \citep[see, e.g.,][]{Shigeyama1988, Hashimoto1989, Thielemann+90, Thielemann1996, Nakamura2001, Nomoto2006, Umeda2008, Moriya2010, Bersten2011, chieffi_pre-supernova_2013, limongi_and_chieffi_2019, Kobayashi+2006, Sukhbold+2016}.

\subsection{A new initial mass remnant mass relation}
\label{sec:imrm}
The initial mass - remnant mass relation (hereafter IM-RM) adopted here has been computed specifically for this work and is provided in Table~\ref{tab:newrel} and visualised in Figure~\ref{fig:comprel}, which will be commented in Sec. 3.4.  In this section we describe in detail how this IM-RM relation is obtained. 
It is obtained by means of hydrodynamical simulations of the explosions of a
subset of models extracted from the database described in detail in \citet[][]{limongi_and_chieffi_2019} and in \citet[][]{RLC24}. In particular, we consider models with initial iron abundance in the range $-3\leq {\rm [Fe/H]} \leq 0.3$ for three values of the initial rotation velocities, namely $ v_{\rm rot}=0,~150,~300{\rm km~s^{-1}} $. Measured values of $ v_{\rm rot}$ for massive stars are sparse, but the largest values are within our maximal value \citep[see e.g.][]{dufton_etal_2006,hunter_etal_2009}. 
The initial composition adopted for the solar metallicity models ([Fe/H]=0, where $\rm [Fe/H]=log(Fe/H)-log(Fe/H)_\odot$) is the one provided by \citet{asplund2009}, which corresponds to a global metallicity of $Z=1.345\times 10^{-2}$. For the sub-solar metallicites we assume the same scaled solar composition with the exception of C, O, Mg, Si, S, Ar, Ca and Ti for which we assume an enhancement with respect to Fe derived from the observations of low-metallicity stars \citep{cayrel2004,spite2005}. As a result of these enhancements, the total metallicity $Z$~corresponding to [Fe/H]=-1,-2,-3 is $\rm Z=3.236\times 10^{-3},~3.236\times 10^{-4},~3.236\times 10^{-5}$, respectively.

These models have been calibrated as follows: (1) the mixing-length value has been set, as usual, to reproduce the present properties of the Sun; (2) the efficiency of the angular momentum transport and rotation driven efficiency have been calibrated to reproduce the surface N abundance observed in the LMC samples of the FLAMES survey that are centered in the clusters NGC2004 and N11 \citep[see][for more details]{limongi_and_chieffi_2019}.
For each pre-supernova model the hydrodynamical calculation is performed by means of the HYPERION code, that solves the full system of hydrodynamic equations coupled with the equation of the radiation transport under the flux-limited diffusion approximation and with the equations describing the temporal variation of the chemical composition due to nuclear reactions \citep[see][for more details]{limongi_and_chieffi_2020}.  

More specifically, to induce the explosion, the inner 0.8 $\rm M_\odot$ of the presupernova model have been excluded from the computational domain and at this mass coordinate a given amount of thermal energy is instantaneously deposited and spread over $\rm 0.1~M_\odot$. Once the energy is deposited a shock wave forms and begins to propagate outward in mass inducing compression, heating and explosive nucleosynthesis. After the initial phase of the explosion, the innermost layers begin to fall back on the initial remnant which progressively increases in mass. Technically, these layers are progressively removed by the computational domain shifting progressively outward the inner boundary. 
At the end of the fallback phase, that in some cases lasts even $\rm \sim 10^4~s$, the final mass cut, i.e. the mass coordinate separating the final remnant from the ejecta, is naturally obtained. The calculations are followed until $\rm \sim 3\cdot 10^7~s$ after the onset of the explosion. This time is long enough to also follow the phase of homologous expansion of the ejected matter and the associated bolometric light curve. In general, for each progenitor star the mass of the remnant depends on the energy that is initially injected, i.e., the higher the initial injected energy the lower the mass of the remnant. In these calculations we calibrate the initial injected energy in order to obtain an explosion energy at infinity of $\rm E_{expl}=10^{51}~erg$ for all the models.
\subsection{Comparison to literature and observational contraints}
\label{sec:complit}
Although this field is very active, and we expect further progress in the next years, to our knowledge the only IM-RM presently available in literature based on pre-supernova models and hydrodynamical simulations, are those provided by \cite{fryer_etal_2012}, \citet{Sukhbold+2016}, and by \citet{limongi_and_chieffi_2019}. The IM-RM relation provided by \cite{fryer_etal_2012} is based on hydrodynamical simulations with a parametrized supernova engine applied to non rotating presupernova models in the mass range $\rm 11-40~M_\odot$ for metallicity corresponding to solar, $10^{-4}$ times solar and zero. The IM-RM relation provided by \citet{Sukhbold+2016} has been obtained by means of 1D hydrodinamical simulations with a calibrated neutrino luminosity (see above) applied only to solar metallicity non rotating presupernova models in the mass range $\rm 12-120~M_\odot$. The IM-RM relation provided by \citet{limongi_and_chieffi_2019} has been obtained by means of 1D hydrodynamical simulations in the framework of the kinetic bomb where the final remnant has been set by assuming that stars with initial mass $\rm M\leq 25~M_\odot$ eject $\rm 0.07~M_\odot$ of $\rm ^{56}Ni$ and that those with initial mass $\rm M > 25~M_\odot$ fully collapse to a black hole. This IM-RM relation covers progenitors stars with initial masses in range $\rm 13-120~M_\odot$, initial metallicities in the range $-3\leq {\rm [Z/H]} \leq 0$ and initial velocities corresponding to $ v_{\rm rot}=0,~150,~300{\rm km~s^{-1}}$.

\begin{figure}
\includegraphics[width=\linewidth]{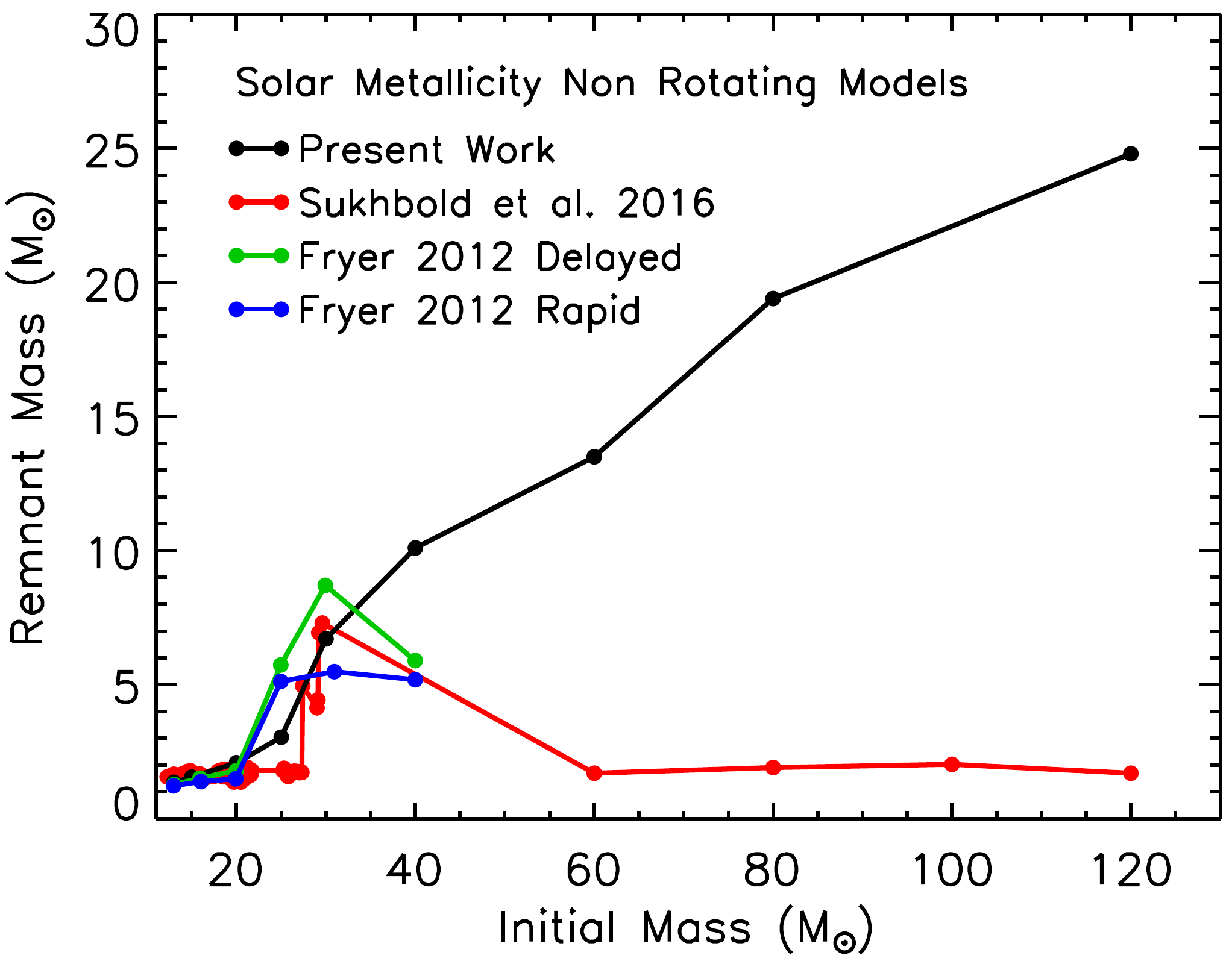}
\caption{Comparison of the initial mass $ M_{\rm in} $ (x-axis) - remnant mass $ M_{\rm R} $ (y-axis) relation obtained in this paper (black line) with those by \citet{Sukhbold+2016} (red line) and by \citet{fryer_etal_2012}, where the green line is for the delayed engine and the blue line for the rapid engine.}
\label{fig:imrmcomparison}
\end{figure}

A comparison between the IM-RM relation obtained here for the solar metallicity non rotating progenitor models and the ones published in \cite{fryer_etal_2012} and \citet{Sukhbold+2016}, i.e. the ones most widely used and based on hydrodynamical simulations and preupernova models (see above), is shown in Figure \ref{fig:imrmcomparison}. All relations are in broad agreement for initial progenitor masses $\rm M\lesssim 35~M_\odot$, taking into account all the possible sources of uncertainties due to the parametrized hydrodynamical simulations and to the differences in the properties of the adopted progenitor models. In particular, it is worth noting the difference between the IM-RM relations obtained by \cite{fryer_etal_2012} for progenitors around $\sim 30~M_\odot$. These relations have been obtained with the same hydrodynamical code and with the same progenitor models, but with two different parametrized supernova engines, i.e., the delayed and the rapid one \cite[see][for more details]{fryer_etal_2012}. The IM-RM relations for stars more massive than $\rm 40~M_\odot$ are only available from \citet{Sukhbold+2016} and the present paper. They differ substantially in the mass interval $\rm 40-120~M_\odot$, which is the one in which the models enter the Wolf-Rayet (WR) phase. 
The discrepancy is due to the assumed mass-loss rate during the WR phase. \citet{Sukhbold+2016} adopt \cite[][LA89]{LA89}, which is about 0.2–0.6 dex higher than the one from \cite[][NL00]{nugislamers2000} adopted by \citet{limongi_and_chieffi_2019} and by the present work. The adopted mass-loss rate affects the CO core masses and $\rm^{12}C$ mass fractions developed after the core He burning, which in turn determine the post-He-burning evolution. For example, a $\rm 60~M_\odot$ star with solar composition computed using NL00 develops a CO core mass of $\rm \sim 12~M_\odot$, whereas the same star modeled with LA89 yields a CO core mass of only $\rm \sim 4~M_\odot$, which will force the stellar evolution to be comparable to that of a $\rm \sim 20~M_\odot$ star \citep[see][for full details]{limongi+2010}.

Our adoption of the NL00 mass-loss rate is supported by the observed WR-to-O star ratio (WR/O) and the WR subtype-to-total WR ratios. The agreement between observed and predicted ratios is generally acceptable for both rotating and non-rotating models, typically within a factor of two. The most significant deviations are found in the WR/O ratio for non-rotating models and in the WNC/WR ratio for the rotating ones. The marked underestimation of the WR/O ratio in the non-rotating scenario suggests that either the threshold initial mass for WR star formation is overestimated or that rotation needs to be included. Moreover, note that this comparison is based on the assumption of a flat initial rotational velocity distribution and a fixed rotational velocity of $\rm 300~{\rm km~s^{-1}}$. A more realistic comparison of number ratios between different WR subtypes and O-type stars should be derived from synthetic stellar population models incorporating a realistic rotation velocity distribution. Such a comparison has not been done yet and is beyond the purposes of the present paper. For further details see \citet{chieffi_pre-supernova_2013}. 

Looking at the luminosities observed for living massive stars in our Galaxy, \citet[][]{sander_etal_2012} report observed WR luminosities in the range $\log(L/L_{\odot})\sim5.1\div5.6$~(see their Figs. 10-11-12). This is in agreement with the luminosity range spanned by our solar-metallicity pre-supernova models (see Figure 14 in LC18), whose luminosities lie in the range $\log(L/L_{\odot})\sim5.3\div6.2$~ for non-rotating models and $\log(L/L_{\odot})\sim5.1\div5.9$~ for models with an initial rotational velocity of $\rm 300~{\rm km~s^{-1}}$.

In summary, comparisons with observations of living massive stars luminosity and number counts support the remnant masses of the most massive stars as in our models shown in Figure~2.

\subsection{Comparison to our previous calculations}
\label{sec:complit}

We now highlight the differences between the new IM-RM relation presented here and the canonical relations we adopted earlier (see Sec. \ref{sec:canrel}). As a main difference, the remnant mass is now naturally obtained as a result of stellar evolution calculations, and depends on the progenitor mass, initial metallicity and initial rotation velocity, while earlier the remnant mass did not depend on the progenitor mass over a large mass interval ($8.5$~to~$40 M_{\odot}$ for NS remnants). The prescription we use here is also an improvement over a previous use of the same pre-SN tracks \citep{mapelli_etal_2020} where the remnant mass was evaluated parametrically based on the compactness of the core without the support of full hydrodynamical calculations.

We further assume that when the remnant mass $ M_{\rm R} \gapprox 2$~$M_{\odot}$, the remnant is a black hole (BH). This implies that BH remnants can form already at $20 M_{\odot}$, or even below depending on metallicity and $ v_{\rm rot}$, instead of the fixed progenitor mass cut of $40 M_{\odot}$ we assumed earlier. The other difference pertains to the mass of the BH remnant, which we previously assumed was half the initial mass (see Sec.~\ref{sec:canrel}). A glance at the table reveals that the canonical assumption for $ M_{\rm R}^{BH} $ is quite close to the results with the new calculations, for low metallicity and high rotational velocity, while it is generally smaller otherwise, namely for high metallicity and for the case of no rotation. 
\begin{figure*}
\includegraphics[width=\linewidth]{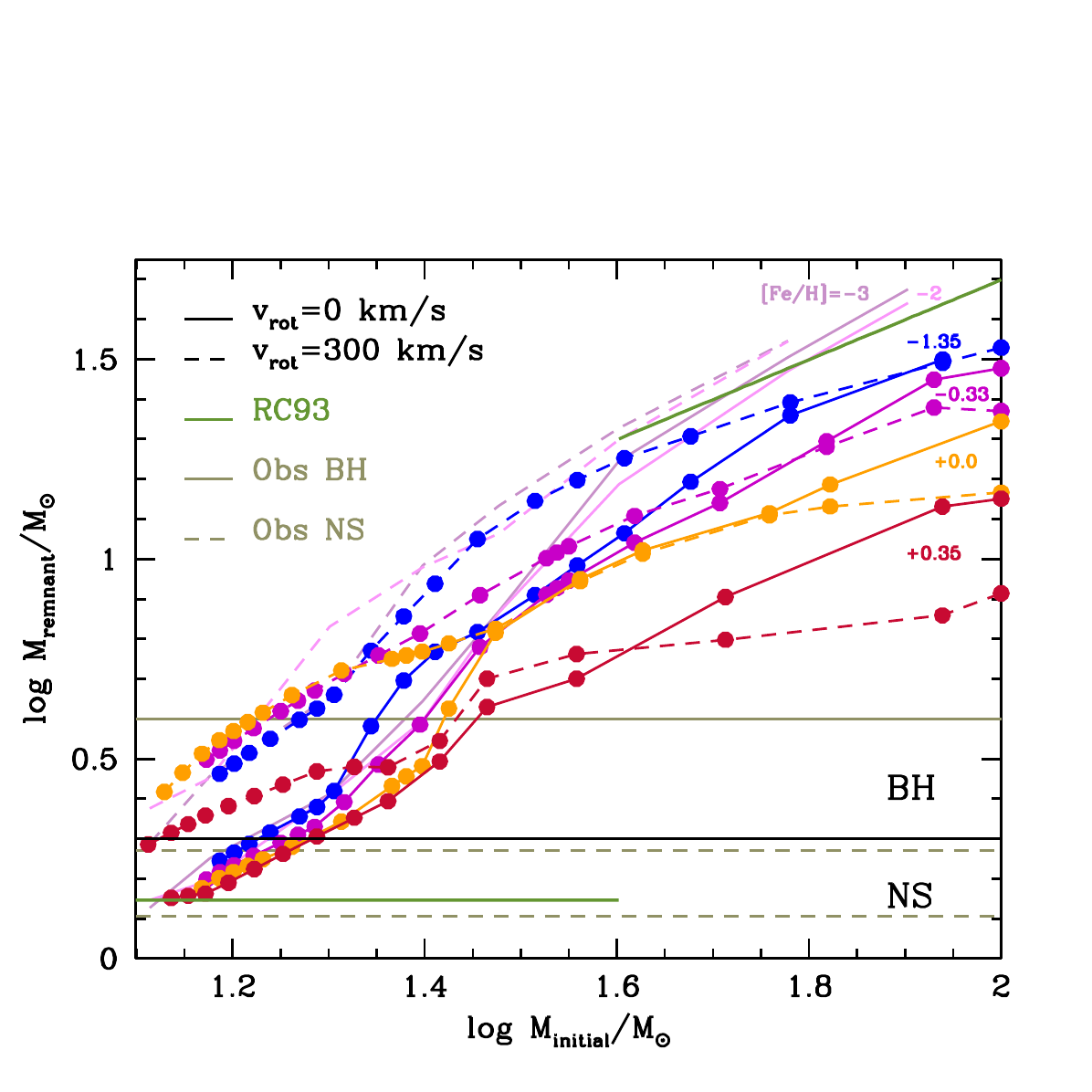}
\caption{Comparison of the initial mass $ M_{\rm in} $ (x-axis) - remnant mass $ M_{\rm R} $ (y-axis) relations obtained with the new calculations described in Sec.~\ref{sec:newrel} (coloured lines labelled by metallicity [Z/H], where values larger than $-2$~already refer to our population model grid, see Sec.\ref{sec:calculations}) and the canonical relations (green solid line). All masses are in units of  $M_{\odot}$. Solid lines refer to the no-rotation case, while dashed lines to a rotational velocity $ v_{\rm rot}=300~{\rm km~s^{-1}} $. Rotational velocity generally implies lighter remnants for $ M_{\rm in} \gapprox 40 M_{\odot}$ and $[Z/H] \gapprox -0.33$~(i.e. half-solar), as an effect of enhanced mass-loss, while at lower metallicity a larger rotation releases a larger remnant, as mostly due to a larger inner burning mass hence final mass. The solid black line marks the 2 $M_{\odot}$ case discriminating NS remnants from BH remnants according to the new relations: at all metallicities a population featured by massive stars with strong rotation velocity may be releasing no neutron stars at all. The upper gray line marks the lower limit of BH masses from GW events \citep[20 $M_{\odot}$,][]{abbott_etal_2020}, while the lower gray line marks the upper limit to observed NS masses reported in the literature \citep[][]{oezel_etal_2012}. Also note that in the latest LIGO/Virgo/KAGRA's third observing run O3 a handful of objects were detected in the $2\div5~M_{\odot}$~mass gap, corresponding to the remnant mass region between the solid gray and black lines.}
  \label{fig:comprel}
\end{figure*}

In summary, the new calculations are consistent with a larger number of generally lighter BH remnants and a marked dependence on metallicity and rotational velocity.

For NS remnants, the range of progenitor mass that now releases an NS is 8.5 -- 15/20 $M_{\odot}$, which is narrower than the range of 8.5 -- 40 $M_{\odot}$ assumed in the canonical relations. Their mass however can be a bit larger than the strict limit of 1.4 $M_{\odot}$ we assumed earlier. Noteworthy, at high rotational velocities the remnant is a BH in all cases, in other words populations featured by massive stars with high rotational velocity do not produce any NS remnant at all above 13 $M_{\odot}$ (i.e. the lower mass limit of these calculations). This result will impact on forecasts for the NS merger background and for statistics of pulsars and X-Ray binaries \citep[][]{pagliaro_etal_2023}.

In summary, the new calculations are consistent with a smaller number of NS remnants and a marked dependence on metallicity and rotational velocity.

Finally, one sees from Table~\ref{tab:newrel} that the highest masses ($ M_{\rm R} \gapprox 80$~$M_{\odot}$) with low metallicity do not produce any remnant at all ($-1$~values in the table), as they most likely conclude their evolution as pair-instability supernovae (PISN) (see, e.g. Limongi and Chieffi 2018). This happens particularly at high rotational velocities. The reason for such a behaviour is the following. The occurrence of the pair instability, which leads to the explosion as PISN, depends on the CO core mass which depends on the He core mass at core- H depletion that, in turn, depends on the initial mass, initial metallicity and initial rotation velocity. In our models, stars that form CO cores larger than $\rm \sim 35~M_\odot$ enters the pair instability regime. Our threshold mass value is consistent with the one found by \citet[][]{heger_and_woosley_2002}. In general, the lower the metallicity, the higher the He core mass at core H depletion (this is also due to the inefficiency of the mass loss at low metallicity), the larger the CO core and therefore the lower the minimum mass exploding as PISN. The effect of rotation is, on the contrary, dual. On one side, stellar rotation facilitates the removal of external layers via mass-loss which leads, for the more massive models, to reduce even the He core, making the CO core less massive, thus increasing the minimum initial mass that enters the pair instability. At the same time, however rotation favours mixing and the inward penetration of fresh fuel during the core H burning, thereby making the He core and therefore also the CO core larger. The outcome of the two competing effects is mass and metallicity dependent and its evaluation requires the detailed calculations we use here.

A visual comparison of canonical and new relations is provided in Figure~\ref{fig:comprel}. 

The remnant mass $ M_{\rm R} $ (y-axis) as a function of the progenitor initial mass $ M_{\rm in} $ (x-axis) is displayed - for the new relations presented in this section - as a function of stellar metallicity (given as Iron abundance [Z/H]) and stellar rotational velocity $ v_{\rm rot}$~(with no rotation as solid lines and $ v_{\rm rot}=300~{\rm km~s^{-1}} $~as dashed lines). The solid black horizontal line delimits the parameter space in terms of either NS ($ M_{\rm R} \lapprox 2$) or BH ($ M_{\rm R} \gapprox 2$) remnants. For comparison, the solid green line shows the $ M_{\rm R} $ vs $ M_{\rm in} $~of the canonical relation by RC93 discussed in Sec.~\ref{sec:canrel}, which forecasts NS remnants from $ M_{\rm in}$ between 8.5 and 40 $M_{\odot}$ and BH remnants for $ M_{\rm in}$ larger than $40 M_{\odot}$ independently of the chemical composition.

In general, a lower chemical composition is consistent with more massive remnants from the most massive stars, i.e. $ M_{\rm in} \gapprox 20 M_{\odot}$, as a result of larger final masses because of a lower mass loss (cfr. Table\ref{tab:newrel}). At $ M_{\rm in} \lapprox 20 M_{\odot}$~the difference in remnant mass is much smaller.

The addition of rotation (dashed lines) generally implies smaller remnants for $ M_{\rm in} \gapprox 40 M_{\odot}$ and high metallicity ($[Z/H] \gapprox -0.33$~(i.e. half-solar)), as an effect of enhanced mass-loss, while at lower metallicity a larger rotation implies a larger remnant, as mostly due to a larger inner burning mass hence final mass. Interestingly, at all metallicities a population featured by massive stars with strong rotation velocity may be releasing no neutron stars above $13 M_{\odot}$~at all according to the new relations, as the remnant is always larger than $2 M_{\odot}$ (we shall return to this point in the Discussion section).

Compared to the canonical relation, as already mentioned, a larger range of initial masses is now concluding their evolution as BHs, namely $ M_{\rm in} \gapprox 13 M_{\odot}$ if $ v_{\rm rot} \neq 0$ and  $ M_{\rm in} \gapprox 20 M_{\odot}$ if $ v_{\rm rot} =0$. These BH remnants cover a wider range of masses, from a minimum of $2 M_{\odot}$ (by construction) to a maximum of $50 M_{\odot}$ for the most metal-poor population with $[Z/H]=-3$, while the minimum BH remnant mass in the canonical relation was~$20 M_{\odot}$. The maximum BH remnant mass was $50 M_{\odot}$, now found only at very low metallicity and in the case of no rotation. It is smaller than $50 M_{\odot}$ in all other cases. Interestingly, the slope of the canonical relation is consistent with the output of the full stellar evolutionary calculations we explore in this paper, particularly in comparison to the solar metallicity case ([Z/H]=0, orange line in Figure~2).

Finally, the gray lines mark current observational constraints to BH and NS masses. Individual BH masses are constrained by current GW detections to be above $\sim~5 M_{\odot}$ \citep[][]{abbott_etal_2020}, and previous constraints from X-Ray binaries made it down to $\sim~4 M_{\odot}$ \citep[][]{corral-santana_etal_2016}. As for NS, the two gray lines emphasize the range of current estimates, i.e. 1.28 to 1.33 $M_{\odot}$ by \citet[][]{oezel_etal_2012}, to which we added the higher value of 1.9 ~$M_{\odot}$~from \citet[][]{nattila_etal_2017}). While these observational limits are probably going to evolve in the next years thanks to the increase number of GW detections, it is nonetheless instructive to show them in comparison to theoretical expectations. Taken at face value, observed NS masses are well explained by our models without rotation, while observed BH masses are explained by a wide range of models with or without rotation and with a range of metallicity. It should finally be emphasized that in the latest LIGO/Virgo/KAGRA's third observing run (O3\footnote{https://observing.docs.ligo.org/plan/}) a handful of objects were detected in the $2\div5~M_{\odot}$~mass gap, corresponding to the remnant mass region between the solid gray and black lines in Figure~\ref{fig:comprel}.

In the next Section we explain how we implemented the new relations into our evolutionary synthesis code.

\begin{table*}
\centering
{\footnotesize 
\begin{tabular}{ l | cccccccc |}
\hline\hline
[Fe/H]& $M_{\rm in}$ & $ v_{\rm rot}=0~{\rm km~s^{-1}} $ & & & $ v_{\rm rot}=300~{\rm km~s^{-1}} $  & \\
 & & $ M_{\rm fin} $ & $M_{Fe}^{\rm core}$ & $ M_{\rm R} $ & $ M_{\rm fin} $ & $ M_{Fe}^{\rm core} $ & $ M_{\rm R} $ \\ \\ \hline \hline
  0.3  &  13 & 11.4  & 1.36 & 1.36  &    4.4      &   1.58 &  1.94      \\         
   &  15  &   11.0 &     1.10 &  1.46   &  5.2 &       1.20 &  2.31      \\
   &  20  &     6.9 &   1.50 & 2.1   &   6.8   &     1.68   & 3.03     \\
   &   25  &    8.2 &  1.45 &  2.73 &   7.5   &       1.25  & 3.00       \\
   &   30  &    9.5 &   1.60 &  4.57 &  8.5     &       1.68 &  5.42        \\
   &   40   &   9.6 &   1.57 & 5.30 &  10.2     &      1.59 &  6.02       \\
   &   60  &   13.9 &    1.56 & 10.0  &10.2      &       1.56 &   6.47        \\
   &   80  &   16.7 &   1.75 & 13.2  &  11.5    &       1.52 &  6.71         \\
 &   120  &   18.7 &  1.48 & 15.1  &   13.6  &        1.65 &  9.70     \\ \hline
 0.0 &   13  & 11.9 &1.36  & 1.36 & 5.4 &              1.63 &  2.38   \\        
      & 15   & 13.2 & 1.43 & 1.55 & 6.2 & 2.77  & 3.39 \\      
     & 20   &  7.5 &  1.1 &  2.09 & 8.2 &  1.94 &  5.19 \\       
     & 25   &  8.5 &  1.38  & 3.04 & 9.5 & 1.94  & 5.87 \\     
     & 30   & 10.8 & 1.57 &  6.71 & 11.2 &  1.65 &  6.74 \\   
     & 40   & 14.1 & 1.53  & 10.1 & 13.8 &  1.91 & 9.93 \\      
     & 60   & 16.9 & 1.52 & 13.5 & 16.6 & 1.76 &  13.3   \\        
     & 80   & 22.7 & 1.66 & 1.94 & 17.5 & 1.77  & 14.0 \\      
    & 120  & 27.9 & 1.92 &  24.8 &  18.6 & 1.72 &  15.3 \\ \hline
-1.0  & 13  & 12.5 &    1.19 &  1.19   &  10.7  & 1.9 &  2.51 \\        
        &  15 &  14.2 &  1.4 & 1.70 & 11.2  & 1.99 &  2.78 \\       
        & 20  & 18.3  & 1.43  & 2.49 & 17.1 &  1.93  & 4.42 \\       
        &  25 &  20.6 & 1.59 & 5.67 & 18.5 & 1.92  & 7.98 \\         
        &  30 &  28.3 & 1.57 & 6.95 & 16.0 & 2.05 & 12.6 \\         
        &  40 &  28.7 & 1.55 & 11.3 & 20.7 & 2.01 & 17.7 \\        
        &  60 & 42.0 & 1.68  & 22.8 & 27.5 &  2.23 &  24.6 \\         
        &  80 & 39.9 &  2.18 & 29.4 &  32.1 & 2.2 &  29.4 \\         
        & 120 &   -1   &     -1    &   -1   & 40.5  &  1.72  & 38.1 \\ \hline
-2.0 &   13 &  13.0 & 1.4 & 1.4 &  11.5   & 1.92  & 2.38  \\        
      & 15  & 14.8 & 1.08  & 1.57 & 13.75 & 2.97 &  2.83 \\       
      & 20  & 19.7 &  1.43 & 2.56 & 2.03 &  6.78 &      16.8   \\     
      & 25  & 24.7 &  1.53 & 3.94 & 1.92  & 9.54 &      13.2   \\     
      & 30  & 29.9 &  1.55 & 7.35 & 1.84  & 11.6 &     15.6   \\    
      & 40 &  39.7 &  1.54 & 15.4 & 2.2  & 19.9 &      22.9    \\     
      & 60  & 59.4 &  1.69 & 29.6 & 1.77 &  35.1 &      37.4  \\       
      & 80  & 78.6 &  1.78 & 43.8 &  47 & -1  &       -1        \\       
      & 120  &  -1   &   -1    &    -1   &    -1       &         -1    &     -1      \\ \hline 
-3.0 &  13  & 13 &  1.15 &  1.34 &  12.8 &  1.17 &  1.94 \\         
      & 15   &  15 &  1.46 & 1.78 &  13.8 &   2.07 &  2.86 \\         
     & 20   & 20 &  1.44 &  2.58 &   20.0 &  1.93 &  4.53 \\               
     & 25    & 25 & 1.53 & 4.41 &  13.3 & 1.89 &  9.65 \\            
     & 30  & 30 & 1.57 & 7.41 &   17.1 & 1.92  & 13.6 \\           
     & 40  & 40 &  1.72 & 17.6 &  24.5 & 2.26 &  21.3 \\           
     & 60  & 60 &  1.75 & 32.0 &  38.1 &  2.53 &  35.2 \\      
     & 80  & 80 &  1.81 & 47.3 &   -1  &  -1  &   -1  &    \\         
    &  120 &  -1 &   -1   &    -1     &    -1   &      -1   &     -1     \\    
 \hline 
 \end{tabular}
}
\caption[Initial final stellar mass relation]{Parameters of the new initial mass-remnant mass relation we explore here. For initial masses of the progenitor star $M_{\rm in}$ from 13 to 120 $M_{\odot}$, and a range of chemical compositions (given as iron abundance [Fe/H]) we list: final masses $ M_{\rm final} $, Iron Fe core masses $M_{Fe}^{\rm core}$ and remnant masses $ M_{\rm R} $, for two value of stellar rotational velocity, namely no rotation $ v_{\rm rot}=0~{\rm km~s^{-1}} $ and $ v_{\rm rot}=300~{\rm km~s^{-1}} $. Values of $-1$ indicate no remnant, due to pair-instability supernovae at high mass and low metallicity. The remnant mass has a strong and non linear dependence on rotational velocity. This acts at reducing the remnant mass by nearly a factor 2 at high mass and high metallicity, but also at guaranteeing a remnant for the highest mass 120 $M_{\odot}$ at low metallicity [Fe/H]=-1, which - with no rotation - leaves no remnant behind (see text for more discussion). Masses are in solar units $M_{\odot}$.}
\label{tab:newrel}
%} 
\end{table*}
\section{Calculation of population remnants}
\label{sec:calculations}
In this Section we describe the steps taken to calculate stellar population models of remnants using the prescriptions discussed in Sec.\ref{sec:newrel}. Firstly, we needed to relate our stellar population model grid - which is expressed in terms of total metallicity [Z/H] and age, the latter being linked to the turnoff mass by its evolutionary timescale (see e.g. Table 1 in M05) - to the new calculations. The first step was to obtain tables of remnant results as in Table~\ref{tab:newrel}, for the metallicity bins of our population models, namely ${\rm [Z/H]}=-1.35,-0.33,0.,0.35$ (M05)\footnote{M05 models also exist for the extreme compositions ${\rm [Z/H]}=-2$ and ${\rm [Z/H]}=+0.67$, but only for ages older than 1 Gyr, whose turnoff mass ($\sim 2 M_{\odot}$) is too small for the production of massive remnants, which are the focus of this paper.}.

As the M05 population models are expressed in terms of total metallicity [Z/H] and are solar-scaled, i.e. [Z/H]=[Fe/H], we calculated the [Z/H] corresponding to the newly adopted stellar evolution by \citet{limongi_and_chieffi_2019} with their assumed [Fe/H]-dependent element enhancement (Section~\ref{sec:newrel}) as [Z/H] = [Fe/H] + $\rm A[\alpha/Fe]$ \citep[][]{thomas_etal_2003}. We found their [Z/H]s are consistent with our [Z/H] grid with the exception of the half-solar model ($\rm{[Z/H]}=-0.33$), which was then obtained by log-interpolation between $\rm{[Fe/H]}=-1.00$ and $\rm{[Z/H]}=0$.  

From now onward we use [Z/H] to indicate total metallicity as this is what defines the M05 population model grid and the galaxy properties (Section~\ref{sec:manga}).

Next, we calculated the remnant masses for our grid of progenitor masses, which we obtained by interpolating in $\log M$~within each metallicity table. 

In our standard population models, the stellar Initial Mass Function (IMF) is populated from 0.1 to 100 $M_{\odot}$. In order to obtain a remnant mass for such a continuous mass range using the discrete grid of Table~1 we adopted 90 $M_{\odot}$~as the maximum mass leaving a remnant and not going through pair instability SN, consistent with the calculations from Limongi \& Chieffi (2018, their Fig. 17). Remnant values for the 90 $M_{\odot}$ critical mass were obtained in log-estrapolation from 80 $M_{\odot}$. For the half-solar model obtained in interpolation, we did not consider the event of pair-instability SN, which usually happens at lower metallicities. Remnant masses for initial masses between 90 and 100 $M_{\odot}$~were obtained by estrapolating the remnant masses of the 80 and 90 $M_{\odot}$ on the $\rm{[Z/H]}=-0.33$ grid.

The total number of remnants of the population $N_{R}$, which is a function of age through the turnoff mass $M_{TO}(t)$, in turn a function of the chemical composition [Z/H], are calculated with standard intergrals, as 
\begin{equation} 
{N_{R} (t,\phi(M),[Z/H])= A\cdot \int_{M_{to}(t,[Z/H])}^{M_{up}(t,[Z/H])} M^{-s} dM}
\end{equation} 
where $\phi(M)=AM^{-s}$~is the adopted IMF function (here for simplicity expressed as a unimodal power-law), with exponent $s$ (in the notation in which the \citet[][]{salpeter_1955} IMF has an exponent of 2.35, i.e. by number) and $A$ is the normalisation constant, such that the total initial mass integrated between the minimum and maximum formed stellar mass $M_{inf}$ and $M_{up}$ (corresponding to $0.1 M_{\odot}$ and $100~M_{\odot}$ in our standard population models) equals 1 $M_{\odot}$. For the same simple case illustrated above, the solution for $A$~is then
\begin{equation} 
A= (2-s)/[M_{up}^{2-s}-M_{inf}^{2-s}]
\end{equation} 
The specific numbers of BH, NS and WD remnants, i.e. $N_{BH}$, $N_{NS}$ and $N_{WD}$ are obtained by integrating over the progenitor mass range that is assumed to leave that remnant type, i.e. for white dwarfs\footnote{$N_{WD}$~is not recalculated here as this paper focuses on the remnants of stars more massive $13 M_{\odot}$. The number of WDs is retained from our previous calculations, see M05 and references therein.}, neutron stars and black holes the following equations hold:
\begin{eqnarray} 
%\begin{flushleft}{ 
N_{\rm WD} (t,\phi(M),[Z/H])= A\int_{M_{TO}}^{8.5 M_{\odot}} M^{-s}dM
\end{eqnarray} 	
\begin{eqnarray} 
N_{\rm NS} (t,\phi(M),[Z/H],v_{\rm rot})=A\int_{8.5 M_{\odot}}^{M_{crit}} M^{-s} dM
\end{eqnarray} 
\begin{eqnarray} 
%\begin{flushleft}{ 
N_{BH} (t,\phi(M),[Z/H],v_{\rm rot})= A\int_{M_{crit}}^{M_{up}} M^{-s}dM
\end{eqnarray} 	
$M_{crit}(t,[Z/H],v_{\rm rot})$ is the initial mass for which the remnant mass equals to $2 M_{\odot}$, which is the adopted threshold to separate NS remnants from BH remnants. For $M_{in}\gapprox M_{crit}(t,[Z/H],v_{\rm rot})$, the remnant is counted as a BH, while for $M_{in}\lapprox M_{crit}([Z/H],v_{\rm rot})$ (and $M_{in}\gapprox 8.5 M_{\odot} $) the remnant is counted as a NS.

Now we turn to the mass budget. The total stellar mass in remnants\footnote{Note that, as we normalise our calculations to $1 M_{\odot}$, the total stellar mass is effectively a fraction that can be conveniently applied to any other analysis.}, hereafter $M^{*}_{R}$, and its partition into BHs, NSs and WDs, hereafter $M^{*}_{WD}$\footnote{As already pointed out for $N^{*}_{WD}$, also $M^{*}_{WD}$~is not recalculated here, see M05 and references therein.}, $M^{*}_{NS}$ and $M^{*}_{BH}$, are obtained via integration by mass as follows. For WDs we have 
\begin{equation} 
M^{*}_{\rm WD} (t,\phi(M),[Z/H])=A\int_{M_{TO}}^{8.5 M_{\odot}} M^{-s}\cdot M_{WD} dM
\end{equation} 
For NS instead,
\begin{eqnarray}
M^{*}_{\rm NS} (t,\phi(M),[Z/H],v_{\rm rot})=
A{\Big(}1.4 \int_{8.5}^{13} M^{-s}dM + \nonumber\\
\int_{13}^{M_{\rm crit}} M^{-s} M_{\rm R}(M_{\rm in}) dM{\Big)}
\end{eqnarray}

In eq.(7) we assign a mass of $1.4 M_{\odot}$~to NS formed in the mass range from $8.5 M_{\odot}$ to $13 M_{\odot}$, that is not covered by the new calculations. This approximation carries little impact on the results because the remnant mass around 13 $M_{\odot}$~is indeed very similar to $1.4 M_{\odot}$ (cfr. Table~\ref{tab:newrel}).
Finally, for BHs the following calculation is performed
\begin{eqnarray} 
M^{*}_{BH} (t,\phi(M),[Z/H],v_{\rm rot})= A\int_{M_{\rm crit}}^{M_{\rm up}} {M^{-s}M_{\rm R}(M_{\rm in} dM}
\end{eqnarray} 	
and finally the total mass stored in remnants at time $t$, $M^{*}_{R}$~is
\begin{eqnarray} 
M^{*}_{R} (t,\phi(M),[Z/H],v_{\rm rot})=M^{*}_{WD}(t,\phi(M),[Z/H]) +\nonumber \\  M^{*}_{NS}(t,\phi(M),[Z/H],v_{\rm rot}) +  M^{*}_{BH}(t,\phi(M),[Z/H],v_{\rm rot}) 
&&  
\end{eqnarray} 

with identical meaning for the various terms as in eqs.3-5. 

As just seen and as well known, the number of, and total mass locked in, stellar remnants depend on the assumed initial mass function (IMF) $\phi(M)$. The effect of the IMF on remnants for stellar population models is comprehensivley discussed in M98 and Courteau et al. (2014), for the models used here. In this work, where the focus is on the new stellar evolution relations for massive stars and remnants, we discuss calculations for one
representative IMF, namely that from \citet[][]{kroupa_2001}\footnote{Models are available for arbitrary options for the Initial Mass Function (IMF)}. This is a widely-used IMF that is empirically-calibrated on Milky-Way star counts. It is the same IMF we used for obtaining the resolved star formation histories of local galaxies within the SDSS-IV/MaNGA project \citep[][]{neumann_etal_2022}, for which we shall use these new calculations in paper II. As a reminder, the Kroupa (2001) IMF is a double powerlaw, with exponents $s1$ and $s2$~of 1.3 and 2.3 (in the notation where the Salpeter's one is 2.35), below and above $0.5 M_{\odot}$, respectively. 

\section{Results}
\label{sec:results}
\subsection{Remnant numbers}
\label{sec:numrem}
\begin{figure}
\includegraphics[width=\linewidth]{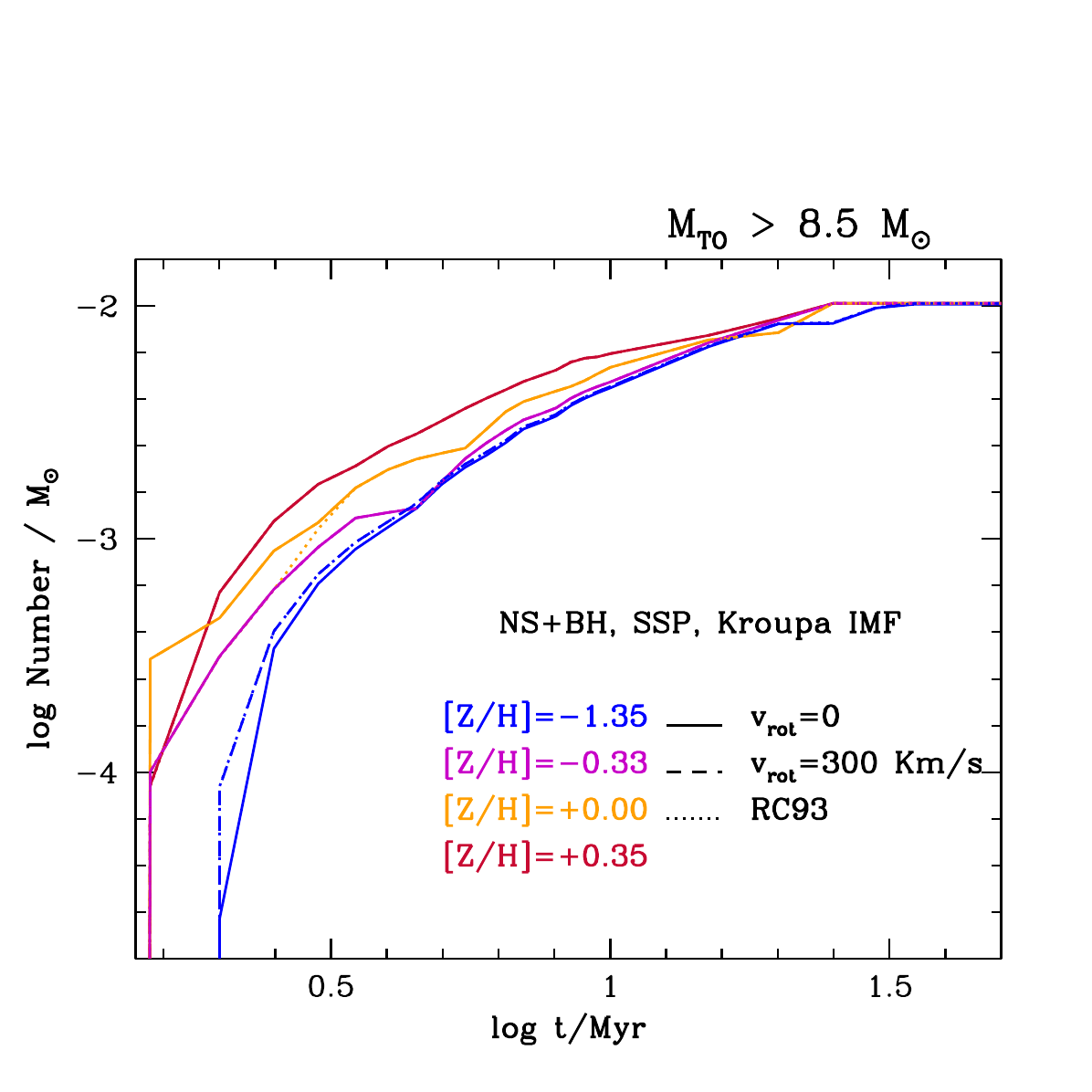}
\caption{Time evolution of the number of massive remnants, $N_{NS}+N_{BH}$, calculated for $1 M_{\odot}$, single-burst populations characterised by a Kroupa IMF and various chemical compositions, labelled with different colours. Solid and dashed lines refer to the $v_{\rm rot}=0$ and $v_{\rm rot}=300 km~s^{-1}$ cases, respectively, while dotted lines to the canonical relations (RC93). There is virtually no difference in the total remnant numbers hence the latter lie underneath the lines for the new IM-RM relations.}
  \label{fig:nt}
\end{figure}

Figure~\ref{fig:nt} show the total number of massive remnants, i.e. $N_{NS}+N_{BH}$, for single-burst stellar population models (i.e. SSPs) with various metallicities (labelled colours). These are zoomed into the relevant ages, i.e. 1 to 50 Myr, corresponding to turnoff masses larger than $\sim 8.5 M_{\odot}$, which undergo supernova and leave this type of remnants. Smaller masses leave a white dwarf (WD) remnant, whose description we have not changed in this paper (see M98, M05, Courteau et al. 2014). In the Figure, solid and dashed lines refer to the new calculations for $v_{\rm rot}=0$ and $v_{\rm rot}=300 km~s^{-1}$, respectively, while dotted lines to the canonical relations (RC93). The latter are barely visible because - as expected from the integrals in Section~\ref{sec:calculations}-
the total number of massive remnants is virtually unchanged with respect to the canonical relations, as the new calculations mostly act at reorganizing the partition of remnants between NS and BH. The only difference appears at the lowest metallicity bin (-1.3, blue line) where - for $v_{\rm rot}=0$ (solid line) and $M_{in}\gapprox 90 M_{\odot}$ - stars go into pair-instability supernovae leaving no remnants. Around 50 Myr and independently of the chemical composition, the total number of massive remnants from any single-burst featured by a Milky-Way-type IMF is just $\sim1$~ per cent the stellar population mass (Figure~\ref{fig:nt}).
\begin{figure*}
\includegraphics[width=\linewidth]{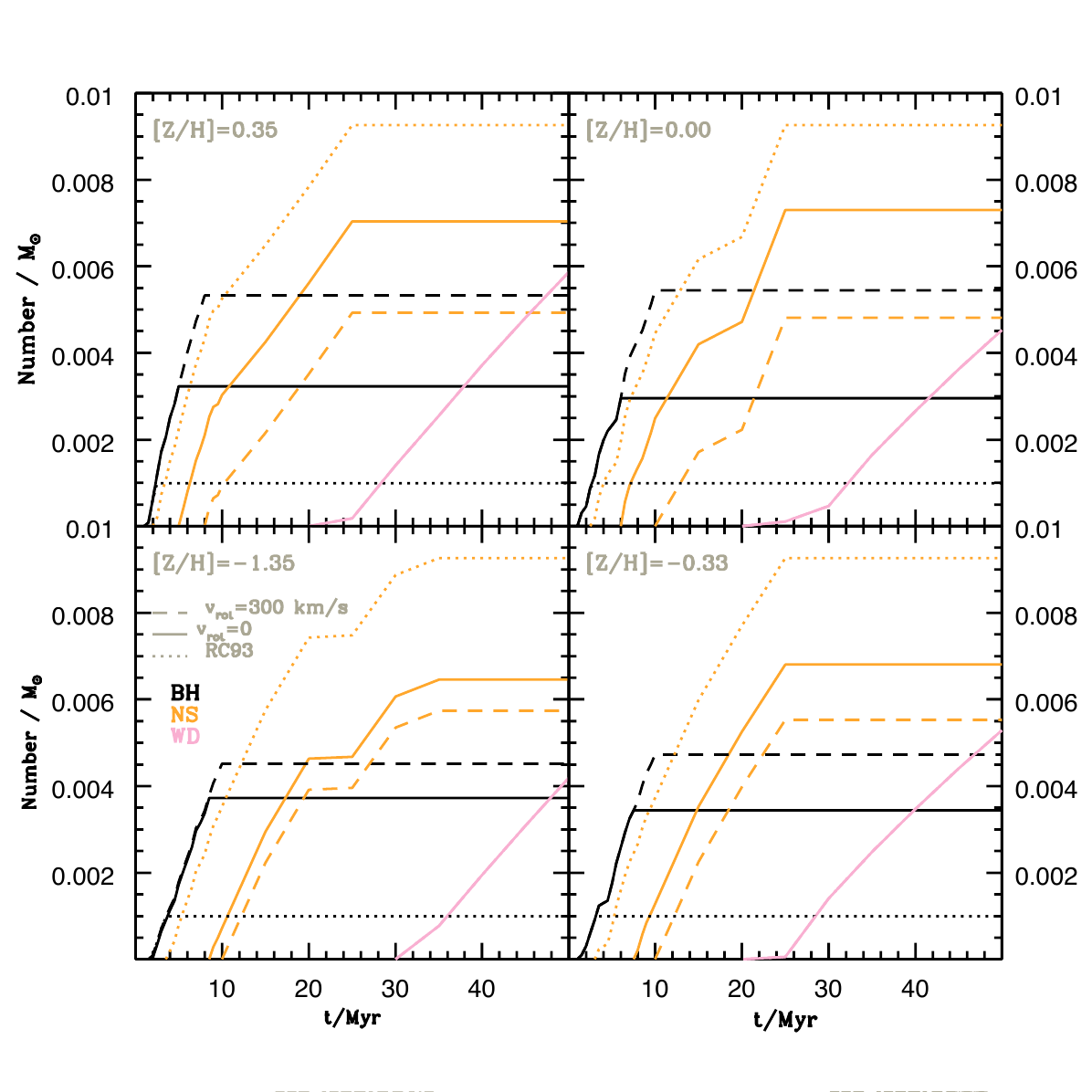}
\caption{Specific time evolution of BH and NS remnants (black and golden lines, respectively) of stellar population models with different chemical composition (one per panel, from the metal-rich double-solar case $[Z/H]=+0.35$ to a metal-poor $1/50~Z_{\odot}$, i.e. [Z/H]=-1.35, from top left to bottom right). Numbers refer to populations with mass equal to $1 M_{\odot}$. The pink solid line shows the time evolution of the number of white dwarf (WD) remnants.}
  \label{fig:nspec}
\end{figure*}
The impact of the new relations is on the specific fractions of NS and BH and on their production epochs. 

These results and their dependence on the various explored parameters are shown in Figure~\ref{fig:nspec}, where each panel refers to a specific chemical composition of the population model, from metal-rich to metal-poor SSPs, from top to bottom and left to right. The time evolution of the number of BH and NS remnants (black and gold lines, respectively), is shown for the case of $v_{\rm rot}=0$ and $v_{\rm rot}=300~km~s^{-1}$ as solid and dashed lines, respectively. The dotted lines depict the results for the canonical relations (RC93). For comparison, the time evolution of the WD remnant component is also shown (pink lines). 

Independently of metallicity and rotational velocity, the number of NSs (orange solid and dashed lines) is lower with respect to the canonical predictions (orange dotted lines) particularly at high metallicity, in favour of a larger number of BHs (black lines, solid and dashed for the new relation, dotted for the canonical relation). A high rotational velocity (dashed lines) boosts the number of BHs, particularly at high metallicity by 0.4 dex, and reduces the number of NS by 0.5 dex around 10 Myr again in metal-rich populations. The onset time of the various remnants is also affected, with rotational velocity acting at delaying the first appearance of NS by a few tenths of Myr (due to the modification of the final mass). The time of the first appearance of BHs in a stellar generation - around just 2 Myr since the formation - is basically invariant - as expected - but the epoch after which the total number of BHs stops growing is pushed forward by $\sim 5$ Myr in models including rotation velocity as a consequence of smaller evolving masses (cfr. Table~1). 
\subsection{Remnant mass budget}
\label{sec:massrem}
\begin{figure}
\includegraphics[width=\linewidth]{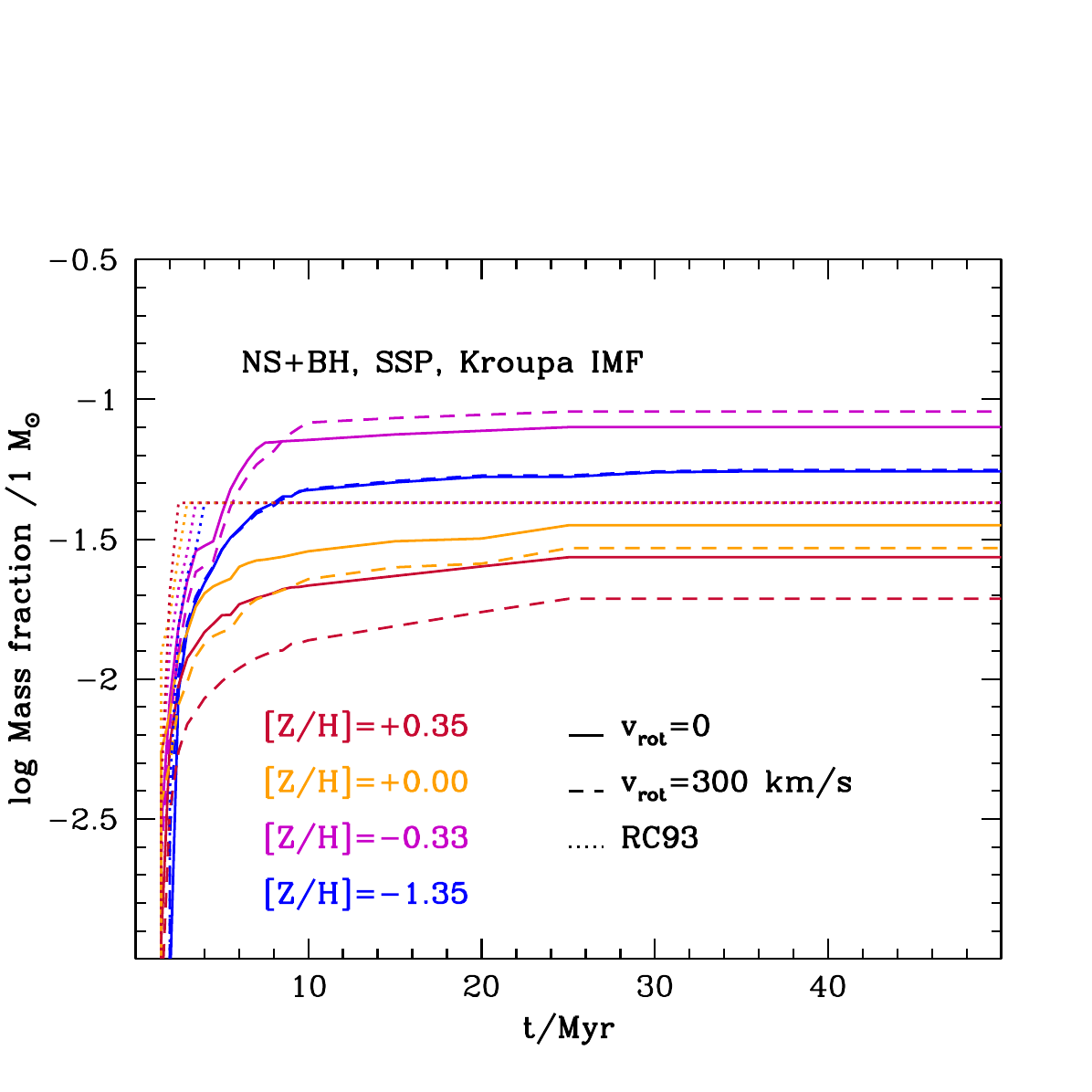}
\caption{Time evolution of the mass fraction locked in massive BH and NS remnants, $M_{{NS}+{BH}}$ (normalised to $1 M_{\odot}$, in single-burst populations characterised by a Kroupa IMF and various chemical compositions, labelled with different colours. Solid and dashed lines refer to the $v_{\rm rot}=0$ and $v_{\rm rot}=300~\rm km~s^{-1}$ cases, respectively, while dotted lines to the canonical relations (RC93). The latter does not depend on the chemical composition, hence dotted lines overlap.}
  \label{fig:mt}
\end{figure}

The mass fraction locked in massive NS and BH remnants - $M_{{NS}+{BH}}$ - for single-burst stellar generations with various chemical composition (coloured lines) and the two assumed extremes for the stellar rotational velocity (i.e. 0 and $300~\rm km~s^{-1}$, solid and dashed lines, respectively) is shown in Figure~\ref{fig:mt}. As usual, the dotted lines show the results for the canonical relations from RC93. As in the latter formulation the remant mass does not depend on the chemical composition (cfr Section 1), the dotted lines overlap.
For the zero rotational velocity case (solid lines), the mass budget in massive remnants of a stellar generation varies from 7 per cent to 1.5 per cent, the lowest value pertaining to metal-rich populations, the highest to metal-poor ones, but not to the most metal-poor one with $[Z/H]=-1.35$~(blue line), due to a combination of remnant mass and turnoff mass, with exact numbers and ranking depending on the assumed IMF. In case of a high stellar rotational velocity (dashed lines), the massive remnant mass contribution further decrease in metal-rich populations (red and orange lines), while remaining essentially the same in subsolar stellar generations. 

\begin{figure*}
\includegraphics[width=\linewidth]{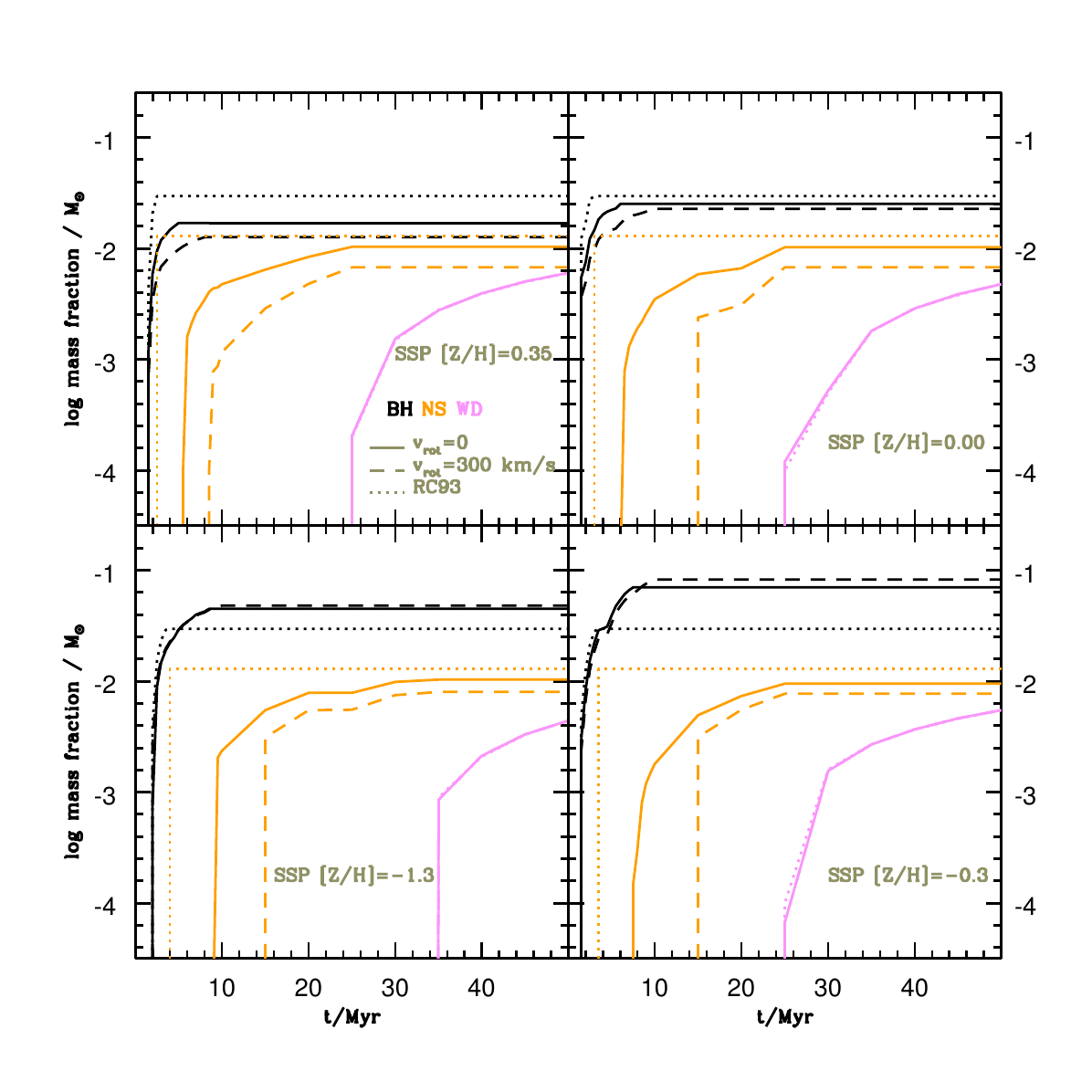}
\caption{Time evolution of the specific mass fractions locked in NS and BH remnants, $M_{{NS}+{BH}}$ (normalised to $1 M_{\odot}$, in single-burst populations characterised by a Kroupa IMF and various chemical compositions, labelled with different colours. Solid and dashed lines refer to the $v_{\rm rot}=0$ and $v_{\rm rot}=300~\rm km~s^{-1}$ cases, respectively, while dotted lines to the canonical relations (RC93).The mass fraction in white dwarf (WD) remnants (pink line) is also shown for comparison.}
  \label{fig:mspec}
\end{figure*}

Similarly to Figure~\ref{fig:nspec}, where we showed the specific numbers of various remnants, we now look at the specific mass fractions from NSs and BHs (plus WDs) in Figure~\ref{fig:mspec}. 
The mass contribution by NSs (orange solid) is lower with respect to the canonical predictions (orange dotted), particularly when stars are rotating (orange dashed). Moreover, as now the NS remnant has a variable mass with respect to the canonical assumption of $M_{NS}=1.4~\rm {M_{\odot}}$, the mass contribution from NSs increases with the stellar population age for then settling onto a constant value of about $\sim2$~per cent around 30 Myr. 

The mass contribution by BHs (black solid) is larger with respect to the canonical predictions (black dotted) at subsolar metallicities and lower at supersolar ones, and is consistent with canonical predictions around solar metallicity. These trends are due to a combination of lower black hole masses $M_{BH}$ with respect to the canonical prediction of $M_{BH}=0.5~\rm {M_{in}}$ because of a wider initial mass range producing BHs stretching down to lower masses (cfr Table 1) and to a higher number contribution from lower masses stemming from the assumed IMF (cfr. Figure~\ref{fig:nspec}). The effect of an initial rotation is mild. A half-solar metallicity population ($[Z/H]=-0.3$) hosts the largest mass contribution from BHs, settling around 8 per cent after 10 Myr.

\begin{figure}
\includegraphics[width=\linewidth]{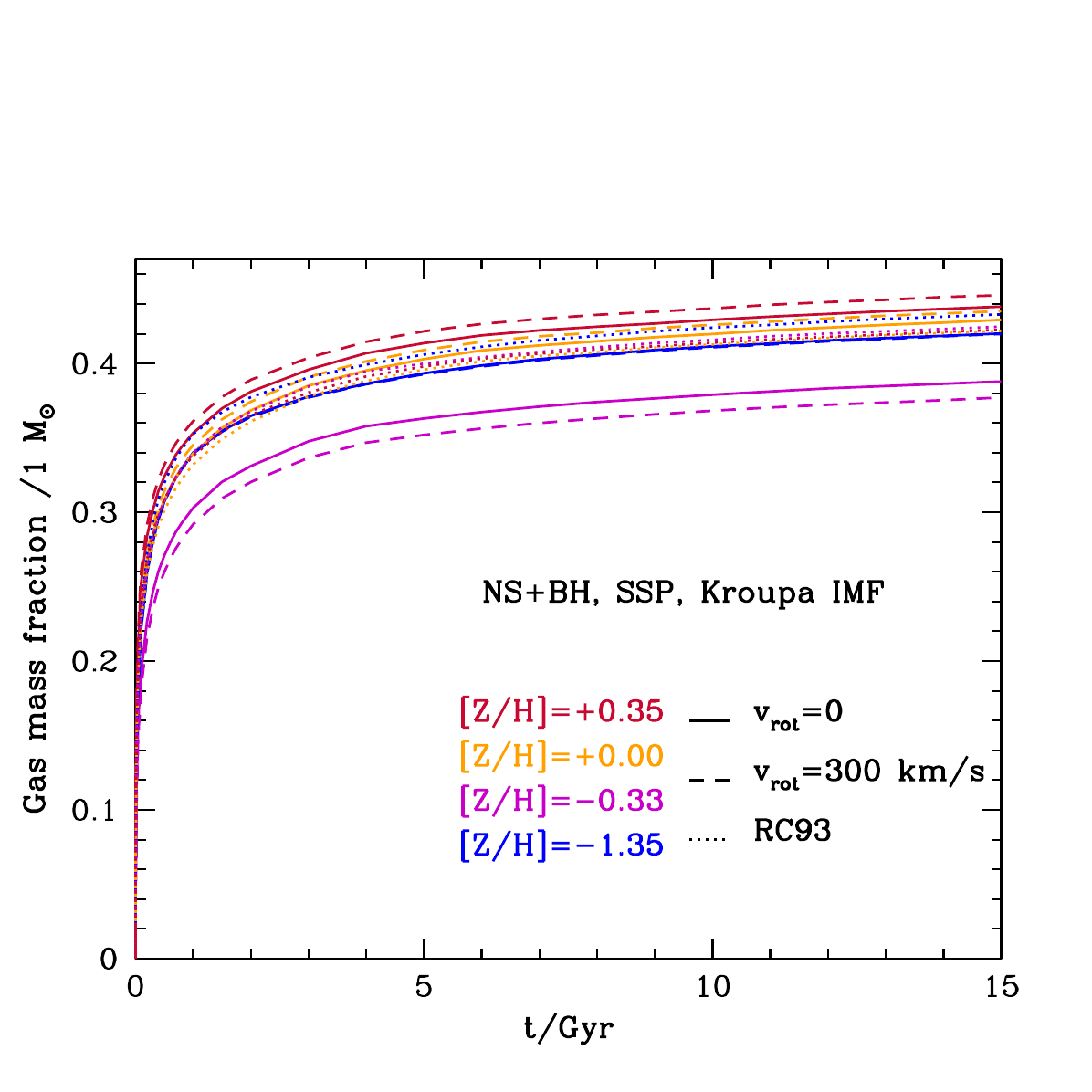}
\caption{Time evolution of the gas mass fraction (normalised to $1 M_{\odot}$), in single-burst populations characterised by a Kroupa IMF and various chemical compositions, labelled with different colours. Solid and dashed lines refer to the $v_{\rm rot}=0$ and $v_{\rm rot}=300~\rm km~s^{-1}$ cases, respectively, while dotted lines to the canonical relations (RC93).}
  \label{fig:mgas}
\end{figure}
The evaluation of stellar mass in living stars and remnants allows the estimation of the ejected gas. The latter can be either recycled into further star formation or ejected out of the system, and is a crucial parameter for chemical evolution models.
Figure~\ref{fig:mgas} shows the time evolution of the gas mass, for the same population models and parameters discussed so far. As the total stellar mass locked in remnants does not change much, so does the gas, settling around 40 per cent in 15 Gyr of evolution. The only noticeable difference is the lower gas mass fraction of the half-solar chemical composition because of the highest mass locked in remnants, as discussed earlier in this section. 
\section{Discussion}
\label{sec:discussion}
The number and mass distributions, and type of stellar remnants after stellar evolution - namely white dwarfs (WDs), neutron stars (NSs) and black holes (BHs) - are critical quantities for galaxy formation, evolution and dynamics, chemical evolution and - when coupled with prescriptions for binary interactions - gravitational wave science. They enter the calculation of the total mass estimated for galaxies hence affecting galaxy dynamics and the determination of any residual mass in dark matter \citep[e.g.][]{beifiori_etal_2014}. They affect chemical evolution by establishing the mass ejecta at supernovae events \citep[e.g.][]{thomas_greggio_bender_1998} and also the availability of NS for more exotic nucleosynthesis \citep[e.g.][]{matteucci_etal_2014,kobayashi_etal_2023}. 

Our standard population modelling \citep[][]{maraston_1998,maraston_2005} adopt a canonical relation for the remnant mass as a function of stellar mass (from Renzini \& Ciotti 1993). This in particular forecasts stars in the mass range 8.5 to 40 $M_{\odot}$~to leave behind a NS of mass 1.4 $M_{\odot}$~ and more massive stars to make a BH with mass half the initial stellar mass, independently of their chemical composition.

In this paper we have revisited our standard population modelling by exploring stellar evolution calculations of the remnant mass as a function the initial stellar mass, for $M \gapprox 13~M_{\odot}$ by \citet[][]{limongi_and_chieffi_2019}, \citet[][]{limongi_and_chieffi_2020} and a bespoke additional set for this work, with a super solar chemical composition, used here for the first time. Also novel is a self-consistent treatment of the supernova evolution, which is performed via hydrodynamical calculations using the Hyperion code, leading to a self-consistent remnant calculation, not requiring extra parametrisation as in \citet[][]{mapelli_etal_2020}. 
The determination of the remnant masses depend on both the progenitor star and the dynamic of the explosion. Here we present for the first time a self consistent determination of remnant masses based on hydrodynamical simulations of the explosion of an extended grid of presupernova models in a wide range of metallicity and for two different rotation velocities. The explosions have been computed in the framework of the thermal bomb approach, which is largely adopted for explosive nucleosyntehsis calculations for core collapse supernovae.

These models forecast a dependence of the remnant mass on the progenitor mass, its rotational velocity and chemical composition, thereby allowing a comprehensive evaluation of the effect from these various parameters. Moreover the remnant type is established by the remnant mass itself, where in particular remnants with mass larger than $2~M_{\odot}$~are counted as BHs. Finally the models also take into account the event of pair-instability supernovae in metal-poor populations, which implies no remnant at all for stellar masses larger than $90~M_{\odot}$. This is in qualitative agreement with independent calculations by \citet{stevenson_etal_2019} who find a maximum BH mass of $40~M_{\odot}$~as a consequence of pair instability supernovae.

With the new relations we have calculated the time evolution of the number and mass contribution of massive stellar remnants in the form of NSs and BHs, for theoretical stellar generations or stellar population models with various chemical compositions spanning from supersolar enrichment typical of massive elliptical galaxies and bulges, to metal-poor environments as those found in dwarf/irregular galaxies or primordial galaxies. We further explore the effect of an initial rotational velocity of the parent star on the stellar population model output. This has interesting and compensating effects, in both augmenting the core mass hence boosting the remnant, but also decreasing the overall stellar mass and speeding up its evolution, thereby affecting the time at which a certain remnant appears in the stellar population number and mass budget. The number and mass of WD remnants is identical to our standard calculations.

Not many works in the literature deal with detailed calculations of remnants for stellar population models. One example is \citet[][]{spera_mapelli_bressan_2015}, who perform calculations linked to the PARSEC stellar tracks \citep[][]{bressan_etal_2012} using new models (at the time) of SN explosions \citep[e.g.][]{fryer_etal_2012}. Similar to our present and earlier works \citep[][]{renzini_and_ciotti_1993,maraston_1998}, they find a BH fraction which is independent of the chemical composition, but heavier BHs at low metallicity , e.g. $60~M_{\odot}$~for a metal-poor population with $[Z/H]\sim-1$. We find the same trend with metallicity (cfr. Figure~2 and Table~1), with a lighter maximum BH of $40~M_{\odot}$~for the same metallicity $[Z/H]\sim-1$. At even lower metallicity, they find BHs as heavy as $130~M_{\odot}$~for $[Z/H]\sim~-2$, while at those low metallicities our calculations - by including Pair-Instability SNs - forecast a maximum BH mass of $\sim80~M_{\odot}$.
Further, they assume a separation between NS and BH at $3~M_{\odot}$, while we assume it at $2~M_{\odot}$. In summary, we find an overall agreement with this previous literature, with some interesting differences that may have some effects on galaxy modelling. For example, we speculate that the absence of super heavy BHs in metal-poor populations affect the $M^{*}/L$~of metal-poor populations such as globular clusters, dwarf galaxies, or primordial galaxies at high-$z$, and may even affect the capability of seeds BH to merge into SMBHs \citep{boco_etal_2021}. 

\begin{figure*}
\includegraphics[width=\linewidth]{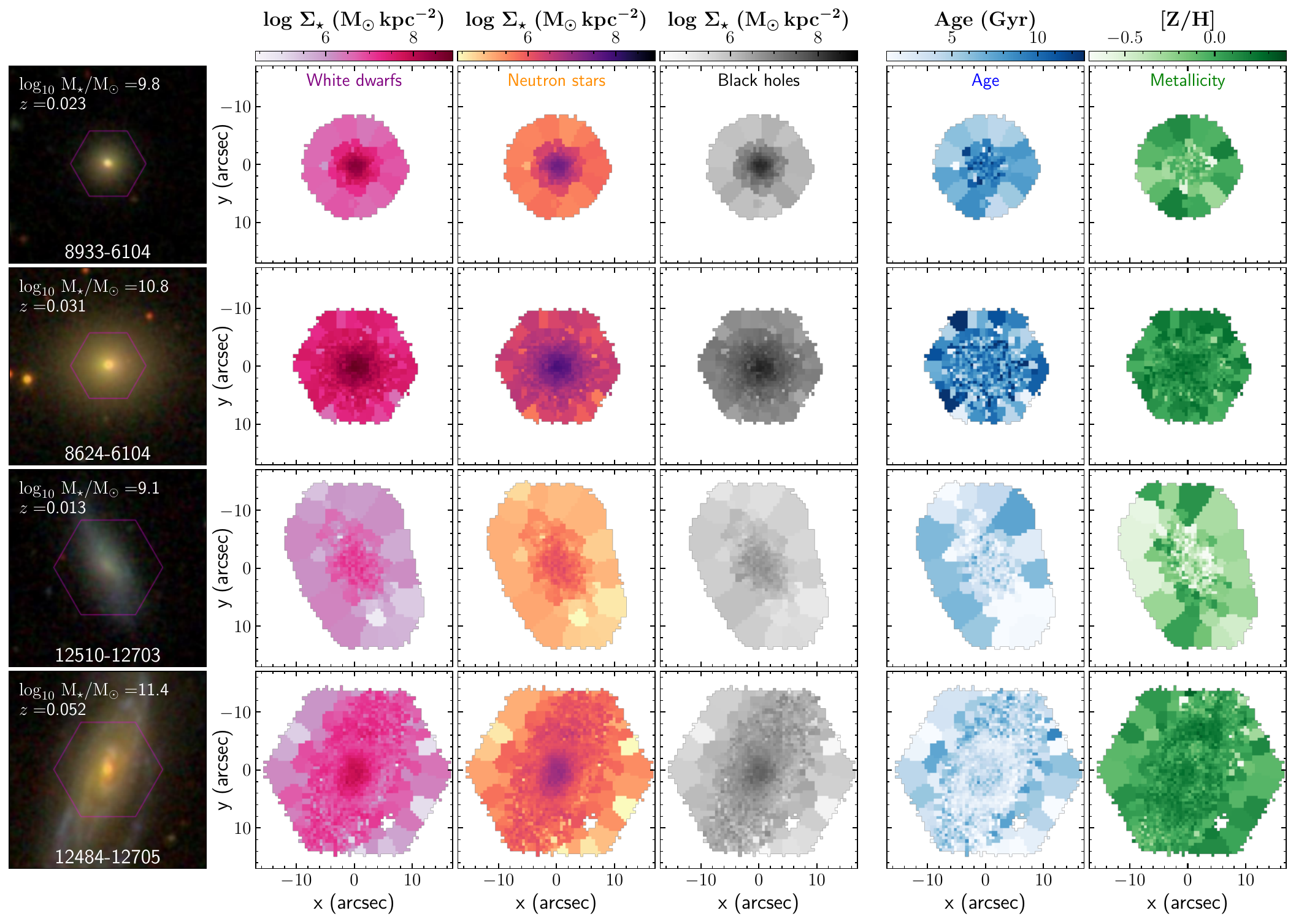}
\caption{Spatially-resolved stellar remnant distributions in galaxies, for illustrative example galaxies (Column 1) spanning a range of type and mass as in N22, Figure~1. Columns 2-4 show remnant mass surface densities, i.e. $M_{R}/kpc^2$. Columns 5-6 report light-weighted age and metallicity maps from N22, for the M11-MILES models, to ease comparisons. These maps are for the $v_{\rm rot}=0~\rm km~s^{-1}$~case.}
  \label{fig:maps_mass}
\end{figure*}

With our models we can now discuss the effect of assumptions on the still uncertain end-point evolution of massive stars. The amount of mass in stars of stellar population models, $M^{*}$, is used ubiquitously in galaxy evolution (e.g. for constraining the driving parameters of galaxy evolution such as mass and environment, as in e.g. \citet{bernardi_etal_2010,thomas_etal_2010,peng_etal_2010}, or the galaxy potentials to constrain dark matter, e.g. \citet{parikh_etal_2024} or for gathering simultaneous information on luminous and dark matter, and the IMF e.g. \citet{smith_lucey_2013}) and cosmology (e.g. for halo occupation distribution and clustering, e.g. \citet{tinker_etal_2010}). Studies of this kind utilise $M^{*}$, which is affected by the amount of remnants included in the population model (see discussion in \citet{beifiori_etal_2011,maraston_etal_2003}). In the following we perform a qualitative comparison of remnant numbers and masses with reference to our canonical relation, as similar ones are adopted in galaxy evolution studies \citep[see][for a comprehensive review on galaxy masses and M/L from various population models]{courteau_etal_2014}. 

As expected, we find that the total number of BH plus NS remnants does not depend on the specific adopted $M_{in}$ vs $M_{R}$~relation, as the minimum mass ($\sim 8.5$) is not affected. There is only a small decrease at low metallicity and young ages due to the occurrence of pair-instability SN implying zero remnants in stars with masses higher than $90~M_{\odot}$. The effect of the new relation is a re-distribution of remnant types, in the sense that there are now more BHs and less NSs than in the canonical relation. Moreover, NSs appear in a stellar generation only at around 10 Myr, or even later (15 Myr) if rotational velocity is included. There are 0.5 dex more stellar mass BHs in galaxies according to the new relations, at all metallicities, and particularly more at high metallicity when considering an initial stellar rotation. The number of NSs and BHs become comparable after $\sim 25$ Myr of evolution. These number statistics could in principle be probed at least qualitatively by gravitational wave detections when the latter will be statistically significant.

Turning now to the mass contribution, the new relations imply a lower total contribution by massive remnants with respect to the canonical relations, by up to 0.6 dex (i.e. a factor 4) at high metallicity. The exception is the half-solar  metallicity, which instead has got a higher mass contribution by 0.2 dex (i.e. a factor 1.6). Taken at face value, this implies lower stellar masses for metal-rich galaxies hence more dark matter, and on the contrary lower dark matter due to a higher stellar mass contribution in mildly metal-poor galaxies. For the metal-poor objects, there is hardly a change. 

The specific mass contributions also change, in particular the new relations imply larger mass stored in BHs with respect to NSs, but - while the mass contribution by NS is always smaller than what is for the canonical relation, the one from BHs has a stronger dependence on metallicity, where for example at solar metallicity the larger numbers and smaller remnant mass balance themselves in such a way as to give very similar mass contributions. A metal-rich population has less mass stored in BHs and a metal-poor one more, the hot spot with the highest mass fraction in BHs being the half-solar metallicity, where as much as 8 per cent of the population mass is in massive BHs.

Before concluding it is worth mentioning that binary evolution which we do not consider may affect the IM-RM relation. For example, binary interactions may modify the IM-FM relation by both allowing for less and more massive remnants due to shifting convection zones during binary interactions \citep[see e.g.][]{eldridge_etal_2017,laplace_etal_2021}. This also implies that NS can appear earlier than the 10~Myr cut off mentioned earlier. Also, the maximum black hole mass may be larger than 40 $M_{\odot}$, depending on which prescription for pair-instability supernovae and binary evolution is employed \citep[see e.g.][but also recent work on the GW transient mass distribution]{stevenson_etal_2019,son_etal_2020,briel_etal_2023}. The inclusion of binary evolution requires the addition of several parameters, while this paper focuses on single-star evolution, whose predictions are relatively clear.

We now turn to the application of our theoretical relations to real galaxies. Indeed, so far we have considered single-burst, single metallicity population models in order to showcase the theoretical trends of building blocks stellar generations.
In the next Section we consider the actual star formation histories of real galaxies in their complexity, contributed by multiple stellar generations with various ages and metallicities, in order to map realistic stellar graveyards in galaxies.
\begin{figure*}
\includegraphics[width=\linewidth]{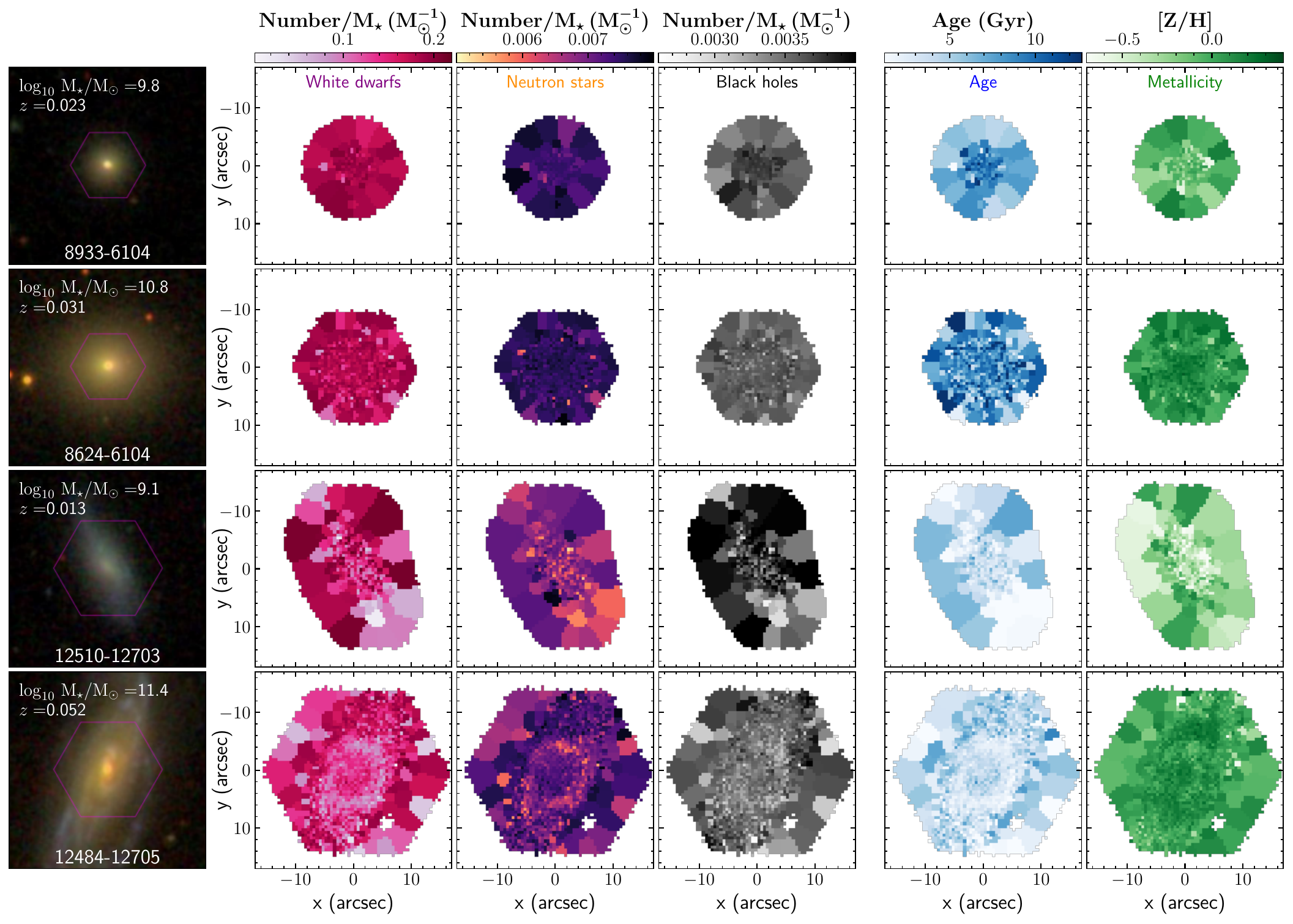}
\caption{As in Figure~\ref{fig:maps_mass} for the remnant number densities, i.e. $N_{R}/kpc^2$. Values are normalised to stellar mass.}
  \label{fig:maps_numbers}
\end{figure*}
\section{Spatially-resolved graveyards in galaxies.}
\label{sec:manga}
The physical properties of galaxies, such as age, chemical composition, star formation history, dust reddening, and stellar mass and mass in remnants, are obtained by comparing theoretical stellar population models to observed spectrophometric data. This kind of analysis is routinely performed in extra-galactic astronomy in order to constrain galaxy formation and evolution. While at high-redshift available data are mostly integrated over the entire galaxy, in the nearby universe it is possible to obtain spatially resolved spectra across single galaxies for a large number of galaxies in order to map their internal structure. As the details of remnants are defined for each stellar population model of given age and chemical composition (see Section~1; Maraston 1998), the stellar population modeling of spatially resolved star formation histories for real galaxies naturally allows the modeling of the spatially resolved remnant population. In this paper we combine the theoretical remnant population presented in this paper, emerging from the IM-FM relations we introduce here, with previously obtained modeled star formation histories for real galaxies in order to map the spatially resolved graveyards in galaxies.  
\subsection{Input galaxy properties}
In this work we use our spatially resolved star formation histories modeled for the MaNGA FIREFLY Value Added Catalog (VAC) (Neumann et al. 2022, hereafter N22\footnote{available at \url{https://www.sdss.org/dr17/manga/manga-data/manga-firefly-value-added-catalog}}). MaNGA (The Mapping Nearby Galaxies at Apache Point Observatory survey, Bundy et al. 2015), a project part of the Sloan Digital Sky Survey-IV (SDSS-IV; Blanton et al. 2017), is the largest Integral Field Unit (IFU) galaxy survey to date containing IFU spectra for 10,010 galaxies in its final data release (SDSS-DR17; Abdurro’uf et al. 2021). Our VAC contains $\sim 3.7$ million spatially resolved stellar population properties - ages and metallicities (both light-weighted and mass-weighted), stellar mass, star formation history, dust reddening, stellar and remnant masses, etc. across 10,010 nearby galaxies encompassing a wide range of morphologies and physical parameters. The full spectral fitting code FIREFLY \citep[][]{wilkinson_etal_2017} is employed to derive the stellar population parameters mentioned above, by fitting linear combinations of single-burst stellar population model spectra to the observed spectra. To model the observed spectra we used two sets of stellar population models, namely M11-MILES \citep[][]{maraston_and_stromback_2011} and MaStar stellar population models \citep[][]{maraston_etal_2020}. Both sets of models are based on the Maraston (1998, 2005) synthesis code, but differ by the assumed stellar spectral library used to distribute the theoretical energetics across the observed wavelength. Although the remnant mass and number per age and metallicity do not depend on the assumed stellar library, the resolved star formation histories do \citep[see][]{neumann_etal_2022}, hence the spatially resolved graveyards as well. In the following, we present results for one set of models, namely the M11-MILES, but we make available also those for the M11-MaStar models\footnote{Results and modeled properties will be available at a dedicated ZENODO page}.

\subsection{Results}
\label{sec:maps}

Figure~\ref{fig:maps_mass} shows remnant maps for four illustrative galaxies that span a range of properties, namely from top to bottom high- and low-mass early- and late-type galaxies, having different star formation histories. We use the same galaxies shown in N22, their Figure~1. It is clear that the mass density of remnants follows the distribution of galaxy stellar mass. This is expected also because we do not include stellar migration in our computations, which may have stirring effects on living stars and remnants specifically in disk galaxies \citep[see e.g.][]{sellwood_and_binney_2002,roskar_etal_2012}. As our star formation histories are derived from the observed light, any past effect of migration is automatically included in living stars, but we cannot assess whether remnants formed earlier in the galaxy history will be located in the same galaxy position now. 

Also interesting for several applications are remnant number distributions in galaxies. Figure~\ref{fig:maps_numbers} is analogue to Figure~\ref{fig:maps_mass}, now showing the number densities of remnants normalised to stellar mass. The distributions are relatively flat across each galaxy, especially in early-type galaxies due to their smooth star formation histories (i.e. smooth distributions of age and metallicity with relatively flat population gradients, see N22 and Goddard et al. 2018). In spiral galaxies, the disk pattern is visible in age and in remnants. Although in these figures remnant numbers are normalized per spaxels, we shall provide absolute numbers later on when discussing the radial profiles of remnants.  

We now turn to examine the effect of the star formation history and stellar properties, namely age, metallicity and stellar rotation, on remnant distributions. In Section~3 we identified the effects of these parameters on the IM-RM relations and here we have the opportunity to study how these effects propagate into galaxy graveyards because of taking into account a galaxy star formation history.  

\begin{figure*}
\includegraphics[width=0.49\textwidth]{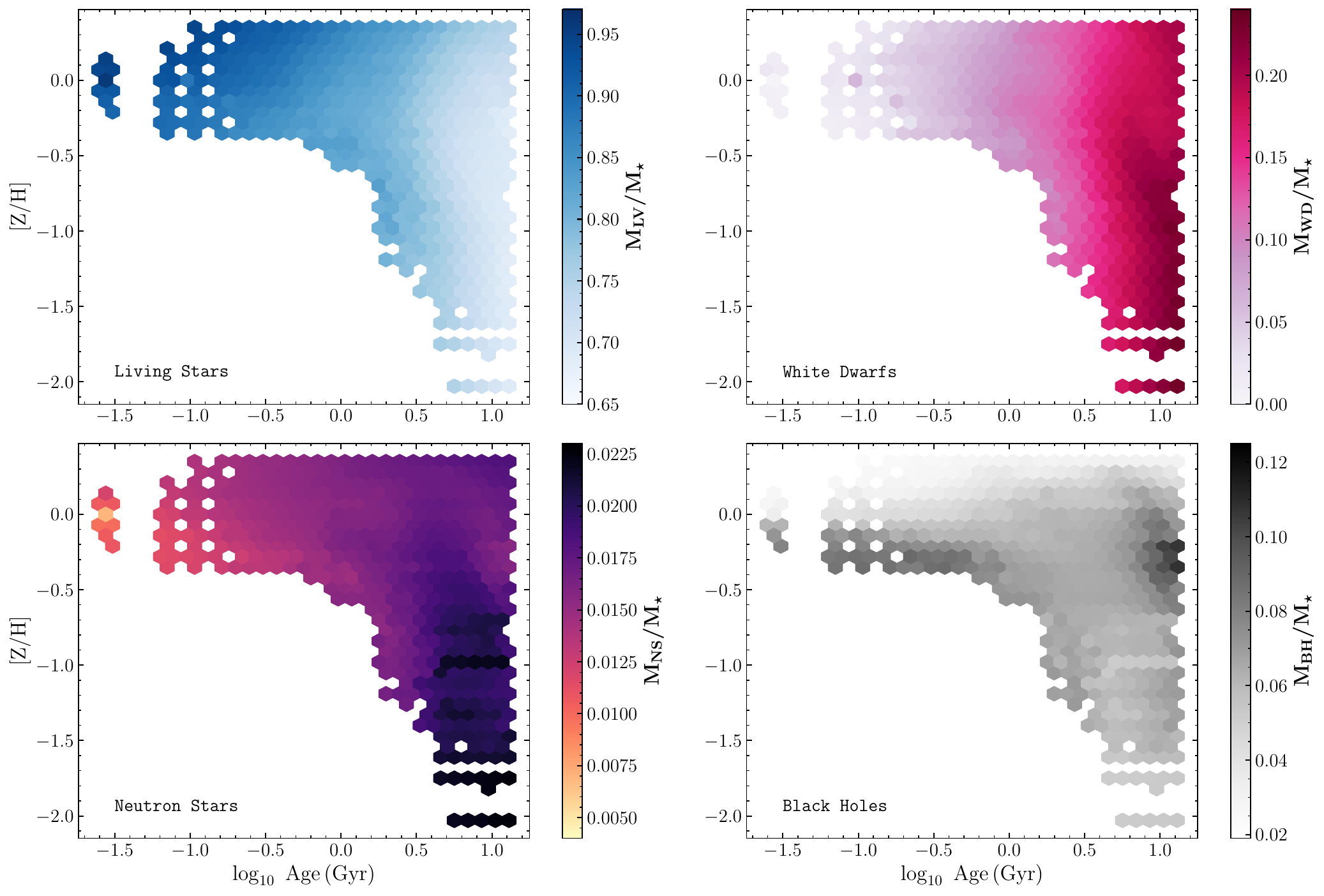}
\includegraphics[width=0.49\textwidth]{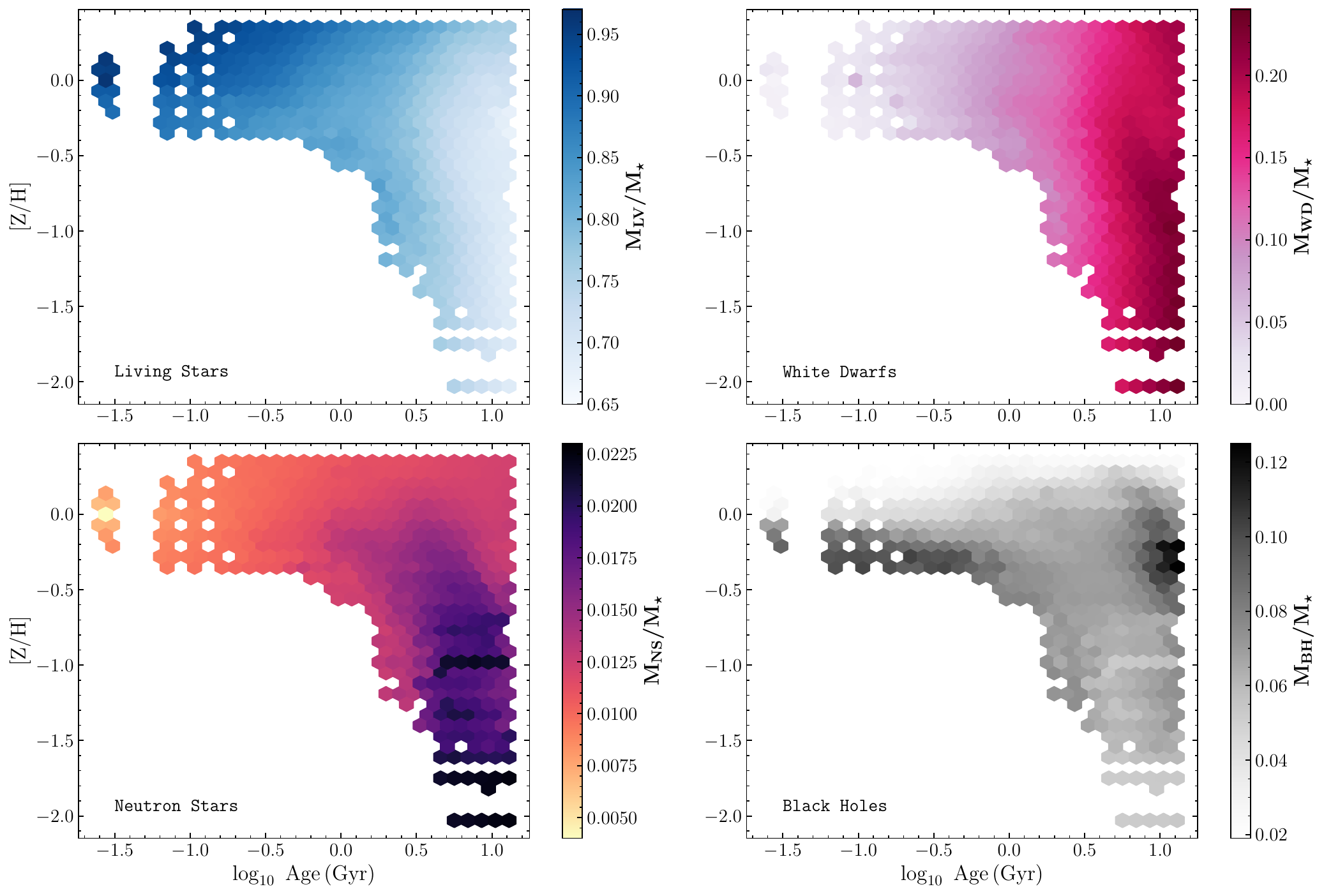}
    \caption{Effect of galaxy age and metallicity on living stars and remnants, for the case of $v_{\rm rot}=0~\rm km~s^{-1}$ and $v_{\rm rot}=300~\rm km~s^{-1}$ (left-hand and right-hand blocks of four panels, respectively). While old age boosts the relative fraction of mass locked in remnants, metallicity leaves a distinct inprint on the mass in BH, which gains a peak at about half-solar metallicity. Stellar rotation reduces the mass contribution by NS in favour of BHs.}
    \label{fig:agezvrot}
\end{figure*}
Figure~\ref{fig:agezvrot} shows the mass contributions from living stars and remnants for $v_{\rm rot}=0~{\rm km~s^{-1}}$ and for $v_{\rm rot}=300~{\rm km~s^{-1}}$ (left-hand and right-hand panels, respectively) as a function of the average age and metallicity of the population, integrated over the whole population of 10,010 local galaxies from the MaNGA survey. Living stars feature young metal-rich populations, whereas remnants have a higher mass contribution in older populations, although the mass contribution by BHs is boosted in young populations as well. Metallicity strongly affects the abundance of BHs, with a peak around half-solar metallicity. These distributions reflect the effect of chemical composition and initial mass on remnant masses (e.g. Fig. 3). For example, the remnant peak at intermediate metallicities ($[Z/H]\sim-0.3$) mirrors the theoretical expectation shown in Figure~6.

The assumption of rotation in massive stars results in a sizable decrease in the number of NSs in favour of BHs, as evident when comparing the bottom panels of Figure~\ref{fig:agezvrot}. This finding may offer an explanation to the missing NSs in the Milky Way over the expectations for Milky Way type-IMFs \citep[see e.g.][]{pagliaro_etal_2023}.

As MaNGA is an IFU survey, we have the opportunity to model the radial profiles of remnants across individual galaxies and examine local as well as global trends. We can further identify analogues of the Milky Way galaxy, where most observations of GW signals come from, in order to disclose any radial trend in remnant distributions, which can be useful for planning observational programmes. MW analogues were sorted following \citet[][]{zhou_etal_2023}, which selects late-type galaxies with masses between $4\times10^{10}~M_{\odot}$ and $8\times10^{10}~M_{\odot}$, a Bulge-to-total light ratio between 0.1 and 0.2 and a $b/a$~ axis ratio larger 0.5.

\begin{figure*}
\includegraphics[width=0.49\textwidth]{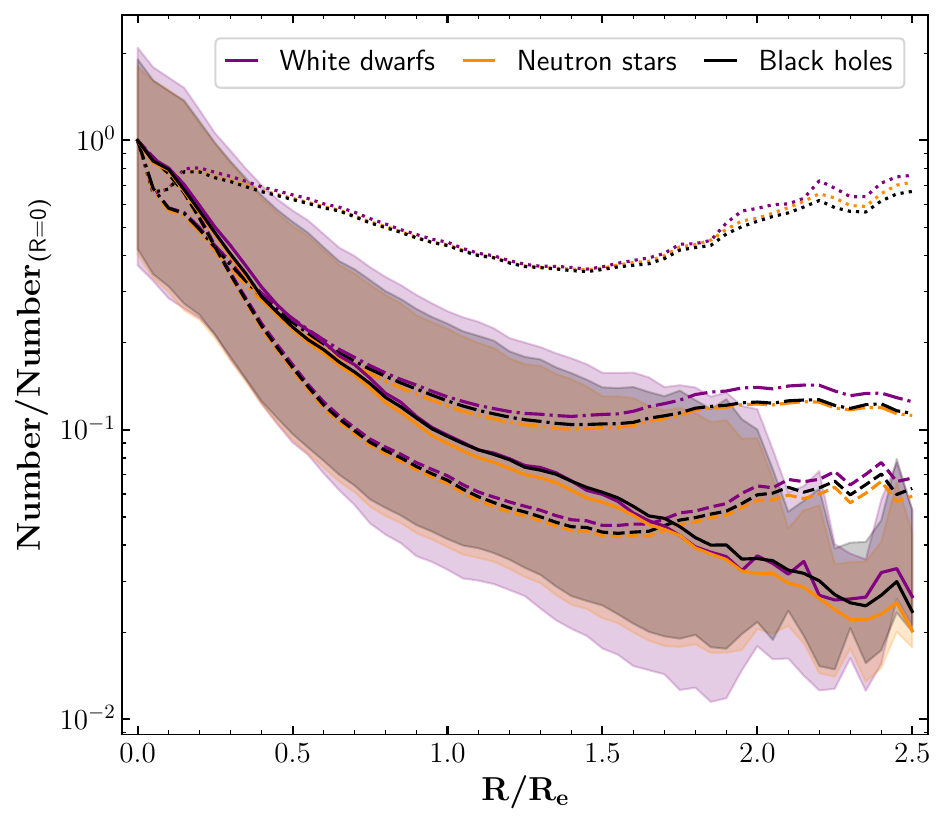}
\includegraphics[width=0.49\textwidth]{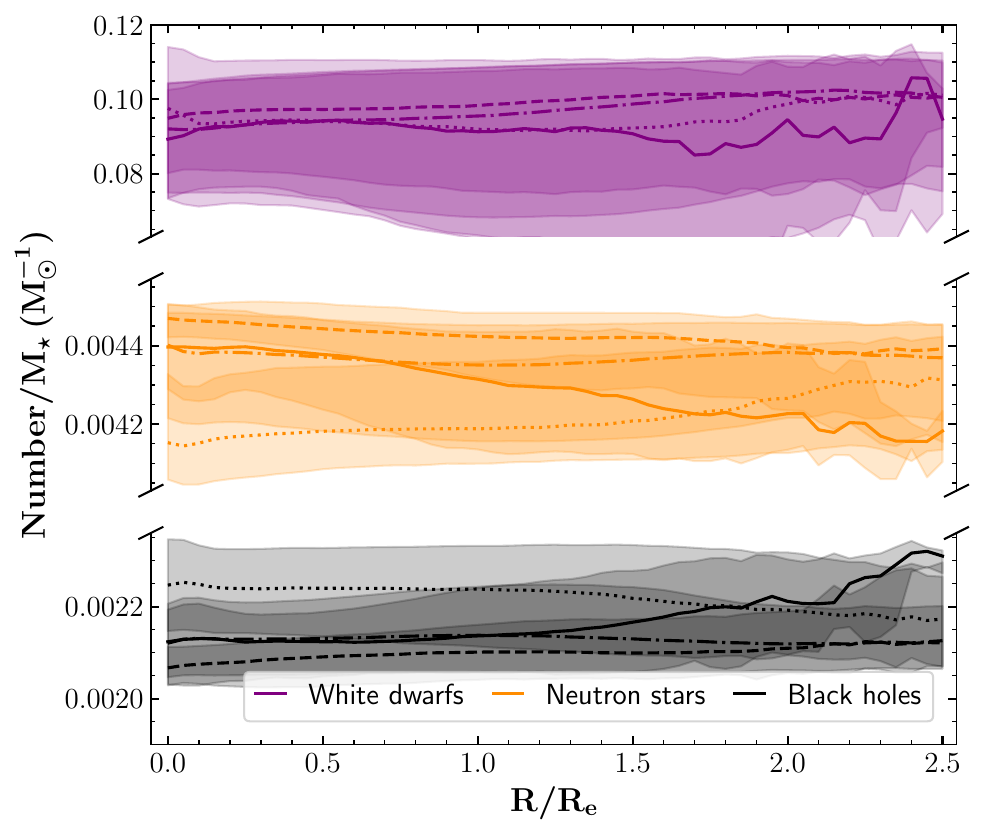}
\caption{{\it Left-hand panel}. Radial profiles of remnants integrated over MaNGA galaxies, split in groups identified by galaxy total stellar mass, as follows: dotted lines refer to low mass galaxies, $M<10^{10}~M_{\odot}$; dot-dashed lines refer to intermediate mass galaxies, $10^{10}~M_{\odot} < M < 10^{11}~M_{\odot}$; 
dashed lines refer to high mass galaxies, $M > 10^{11}~M_{\odot}$; solid lines are for Milky Way analogues, according to the selection by \citet[][]{zhou_etal_2023} (see text). Numbers are normalised to the central values at R=0 to identify gradients.{\it Right-hand panel.} Radial profiles of absolute remnant numbers per stellar mass, integrated over MaNGA galaxy types as defined by various linestyles as in the left-hand panel.}
\label{fig:gradients}
\end{figure*}
Figure~\ref{fig:gradients} provides the radial profiles of remnants from the galaxy center to 2.5 time the galaxy effective radius ($R_{e}$), normalised to the central values in the left-hand panel and in absolute numbers normalised to stellar mass in the right-hand panel. For probing any dependence on galaxy mass, which is one of the main drivers of galaxy evolution \citep[e.g.][]{thomas_etal_2010}, we split the MaNGA sample by stellar mass in bins running from low-mass to high-mass galaxies, as defined in the caption through various linestyles. Solid lines show the results for Milky Way analogues.

We find that remnant numbers display a gradient in intermediate to high mass galaxies, followed by a flatter profile at $R>1 R_{\rm eff}$, while remnant gradients are flat in low-mass galaxies. As these gradients mirror the mass distribution in galaxies, this implies a uniform, rather than concentrated mass distribution in low-mass galaxies. Gradients originate from compensating effects of stellar mass and old age boosting remnants and lower metallicity zones favouring remnants, as stars in metal-poor populations have more massive remnants (cfr. Figure~3). 

The right-hand panel of Figure~\ref{fig:gradients} provides absolute numbers normalised to the local stellar mass (in the spaxels). The effect at finding mostly flat trends is due to the normalisation as spaxels in external regions contain lower stellar mass. The radial distribution of absolute numbers can be scaled to masses of arbitrary galaxy regions to obtain estimates of the expected remnant numbers. These will be upper limits as we do not include the effect of binary stars in our models. Despite the fact that massive stars are likely born in multiple systems \citep[][]{sana_etal_2012}, however the theoretical prediction is that most black holes will end up as isolated due to processes such as supernova kicks, mass loss, or stellar mergers \citep[][and references therein]{mapelli_etal_2020}. 
\section{Summary and Conclusions}
\label{sec:summary}
This paper has a dual scope. The first part presents an update of our previously published stellar population models of remnants, i.e. white dwarfs, neutron stars and black-holes \citep[][]{maraston_1998,maraston_2005}, using a new initial mass - final mass relation for massive remnants neutron stars and black-holes obtained from 1D hydrodynamical simulations based on pre-supernovae models including a wide range of parameters such as initial mass, chemical composition and rotational velocity \citep[][]{limongi_and_chieffi_2019,limongi_and_chieffi_2020,RLC24}. Originally for this paper, these relations have been calculated for a super-solar metallicity, which is known to feature massive galaxies and bulges \citep[e.g.][]{thomas_etal_2010}. The initial mass - final mass relation for white dwarfs is kept from our standard models. We compare the new initial vs final mass relations to literature calculations, finding larger remnants for $M_{\rm initial}> 40 M_{\odot}$~mostly due to a lower stellar mass loss in our stellar evolution models. We then applied the new relations to our evolutionary population synthesis and derive the time evolution of number of remnants and mass locked in remnants, for single generation stellar population models spanning a wide range of ages and chemical composition. Furthermore, we explore the effect of stellar rotational velocity on stellar population models of remnants, for two different assumptions, namely, no rotation and a maximal rotation of 300 $\rm km~s^{-1}$. We find that the main effect of stellar rotation is suppressing the formation of neutron stars in favour of black-holes, mostly due to rotation causing inward penetration of fuel and larger stellar cores. This result may have interesting implications for the missing NS case discussed in the literature \citep[e.g.][and references therein]{pagliaro_etal_2023}. We also find, consistently with previous work, that metal-rich stellar generations give origin to smaller remnants. Here we additionally find that a half-solar metallicity seems to be consistent with the highest number of remnants, as a consequence of the metallicity dependence in the initial-mass - final-mass relation. As an original application of our new models, we then coupled the remnant population models with previously obtained resolved star formation histories \citep[][]{neumann_etal_2022} of a large population of local galaxies from the MaNGA IFU survey \citep[][]{bundy_etal_2016}. This allows us to probe spatially resolved graveyards in galaxies as a function of galaxy mass, star formation history, age, chemical composition, and stellar rotation. We find that galaxy age and metallicity affect the number of and mass in, remnants, with more metal-rich galaxies hosting fewer remnants, although the higher stellar mass of metal-rich galaxies partly compensates for this effect. We also find that radial gradients in the number of remnants depend on galaxy mass mostly because of the mass-dependent profiles in mass density: gradients are flat in low-mass galaxies and negative in high-mass galaxies, particularly in Milky Way analogues. 

To conclude, our predictions rely on a number of assumptions, from the mass loss rate in the assumed stellar models (Section~\ref{sec:newrel}) to the stellar population models and fitting approach utilised to infer the galaxy star formation histories (Section~\ref{sec:manga}). In particular, the absolute numbers and relative fractions of NS vs. BH depend on the adopted stellar mass loss rate. As we are dealing with dark remnants, testing our predictions with observations is not trivial. However, follow-up of GW transients should provide clues into the radial distribution of remnants. For example, \citet[][]{levan_etal_2017} could constrain the physical location of a binary NS merger in a lenticular galaxy. As more transients are discovered their location within host galaxies will allow the build-up of a constraint on the distribution of remnants within galaxies that we predict in this paper.
\section*{Acknowledgments}
CM acknowledges discussions with Maria Alessandra Papa and the hospitality of the Einstein Institute where the foundations of this work were first discussed. The authors thank Ian Harry for discussions on observational limits for NSs and BHs and Alfonso Aragon-Salamanca for discussions on an initial version of the project. Finally, we acknowledge a constructive report from an anonymous referee that helped us improve and expand the article.

\section*{Data Availability}
The models and modeled data underlying this article, i.e. the initial mass - final mass relation, the stellar population models of remnants and the modelled graveyards of MANGA galaxies are publicly available via Zenodo at \url{https://zenodo.org/records/15397763}.
%
%%%%%%%%%%%%%%%%%%%%%%%%%%%%%%%%%%%%%%%%%%%%%%%%%%



\bibliographystyle{mnras}
\bibliography{remnants} % if your bibtex file is called example.bib

%%%%%%%%%%%%%%%%%%%%%%%%%%%%%%%%%%%%%%%%%%%%%%%%%%

%%%%%%%%%%%%%%%%% APPENDICES %%%%%%%%%%%%%%%%%%%%%

\appendix
%%%%%%%%%%%%%%%%%%%%%%%%%%%%%%%%%%%%%%%%%%%%%%%%%%

% Don't change these lines
\bsp	% typesetting comment
\label{lastpage}
\end{document}